\documentclass[preprint2]{aastex}

\slugcomment{Not to appear in Nonlearned J., 45.}

\shorttitle{Spitzer-IRAC Emission in Herbig-Haro Objects}
\shortauthors{Takami et al.}

\begin{document}

\bibliographystyle{astron}


\title{A Detailed Study of Spitzer-IRAC Emission in Herbig-Haro Objects (I): Morphology and Flux Ratios of Shocked Emission}


\author{Michihiro Takami\altaffilmark{1}, Jennifer L. Karr\altaffilmark{1}, Haegon Koh\altaffilmark{2}, How-Huan Chen\altaffilmark{3,1},
Hsu-Tai Lee\altaffilmark{1}}

\altaffiltext{1}{Institute of Astronomy and Astrophysics, Academia Sinica.
P.O. Box 23-141, Taipei 10617, Taiwan, R.O.C.; hiro@asiaa.sinica.edu.tw}
\altaffiltext{2}{Astronomy \& Space Science Department, Sejong University, 98 Kwangjin-gu, Kunja-dong, 143-747, Seoul, South Korea}
\altaffiltext{3}{Department of Physics, National Tsing Hua University, 101 Section 2 Kuang Fu Road, Hsinchu, Taiwan 30013, R. O. C.}


\begin{abstract}
We present a detailed analysis of Spitzer-IRAC images obtained toward six Herbig-Haro objects (HH 54/211/212, L 1157/1448, BHR 71).
Our analysis includes: (1)  comparisons in morphology between the four IRAC bands (3.6, 4.5, 5.8 and 8.0 \micron), and H$_2$ 1-0 S(1) at 2.12 \micron~ for three out of six objects; (2) measurements of spectral energy distributions (SEDs) at selected positions; and (3) comparisons of these results with calculations of thermal H$_2$ emission at LTE (207 lines in four bands) and non-LTE (32--45 lines, depending on particle for collisions). We show that the morphologies observed at 3.6 and 4.5 \micron~ are similar to each other, and to H$_2$ 1-0 S(1). This is well explained by thermal H$_2$ emission at non-LTE if the dissociation rate is significantly larger than 0.002--0.02, allowing thermal collisions to be dominated by atomic hydrogen. In contrast, the 5.8 and 8.0 \micron~ emission shows different morphologies from the others in some regions. This emission appears to be more enhanced at the wakes in bow shocks, or less enhanced in patchy structures in the jet. These tendencies are explained by the fact that thermal H$_2$ emission in the 5.8 and 8.0 \micron~ band is enhanced in regions at lower densities and temperatures. Throughout, the observed similarities and differences in morphology between four bands and 1-0 S(1) are well explained by thermal H$_2$ emission.
The observed SEDs are categorized into:-  (A) those in which the flux monotonically increases with wavelength; and (B) those with excess emission at 4.5-\micron. The type-A SEDs are explained by thermal H$_2$ emission, in particular with simple shock models with a power-law cooling function ($\Lambda \propto T^s$). Our calculations suggest that the type-B SEDs require extra contaminating emission in the 4.5-\micron~ band. The CO vibrational emission is the most promising candidate, and the other contaminants discussed to date (H I, [Fe II], fluorescent H$_2$, PAH) are not likely to explain the observed SEDs.
\end{abstract}





\section{Introduction}
Infrared Array Camera (IRAC) on the Spitzer Space Telescope has been extensively used for studying young stellar objects (YSOs).
The camera has shown that active YSOs and star forming regions are associated with extended infrared emission,  which is in many cases attributed to either shocks \citep[e.g.,][]{Noriega-Crespo04, SmithH06, Walawender06, Teixeira08, Neufeld09, Ybarra09}, scattered continuum
in the outflow cavity \citep[e.g.,][]{Tobin07, Tobin08}, or PAH emission \citep[e.g.,][]{Qiu08}. Many authors have made such identification
using three-color images, assigning blue, green, and red to 3.6, 4.5, and 8.0 \micron, respectively (see the above references). 
This method is not quantitative, since in principle, individual authors adjust the brightness of each color arbitrarily. 
Even so, stars often appear ``blue'' in such images due to the fact that the flux is larger at shorter wavelengths, while PAH emission appears ``red'' due to excessive emission at 8.0 \micron. As a result, the remaining extended emission often appears ``green''. 

The nature of extended infrared emission in Spitzer IRAC images has not been fully investigated, in particular for this extended  ``green'' emission. Shocks often appear green in three-color
images, and it is often explained by H$_2$ emission due to the morphological similarity to H$_2$ 1-0 S(1) at 2.12 \micron~ 
\citep[e.g.,][]{Walawender06, Tobin07}, and 
some analysis with flux ratios \citep{Neufeld09, Ybarra09}. In contrast,
a number of authors suggest possible contribution in shocked emission from additional lines and bands, including CO, H I, [Ar II], [Fe II], fluorescent H$_2$, and PAH emissions \citep{Noriega-Crespo04, Reach06, SmithH06, Neufeld09, Ybarra09}. Furthermore, scattered continuum in the outflow cavity also tends to appear ``green" in three-color images \citep[e.g.,][]{Tobin08}. 
Such an uncertainty has led to the emergence of a new category of YSOs, namely ``extended green objects'' \citep{Cyganowski08}, whose nature is being investigated \citep{Cyganowski09}. Better understanding of this ``green'' emission would allow further understanding for shock conditions and activities for a number of YSOs observed using IRAC.

To investigate the nature of shocked emission observed using IRAC, \citet{Neufeld09} observed spectra in outflows at 5--37 \micron~ and discussed emission in the four IRAC bands based on their calculations of simplified shock models. \citet[][hearafter YL09]{Ybarra09} have made preliminary calculations of molecular hydrogen in non-LTE with isothermal slabs with a larger number of transitions, and perform comparisons with observations of HH 54. Following these studies, we present a detailed investigation of Spitzer-IRAC emission in shocks associated with six Herbig-Haro objects (HH 54/211/212, L 1157/1448, BHR 71). 
Our research goal is to, with the data set available at present, fully investigate the nature of this shocked emission: i.e., whether they are explained by thermal H$_2$ emission; the implications for temperature (or temperature structures) and density, and the degree of thermalization (LTE or non-LTE) of these regions; and possible contaminating emission.
The above outflows contain a variety of structures such as bow shocks, knotty structures, and a bubble-like structure, and are thereby ideal for investigating emission in shocks for a variety of conditions.

We investigate morphological similarities and differences between the four IRAC bands (3.6, 4.5, 5.8 and 8.0 \micron), and H$_2$ 1-0 S(1) for three out of six objects obtained using data obtained by Subaru-MOIRCS and VLT-ISAAC. We also classify spectral energy distributions (SEDs) observed at selected positions. These results are compared with:-  (1) LTE calculations of thermal H$_2$ with $\sim200$ lines; and (2) non-LTE calculations of thermal H$_2$ emission with 32--45 lines for collisions of  H+He and H$_2$+He.
The remaining part of the paper is organized as follows.
In \S 2 we describe data we have used, including new observations using Subaru-MOIRCS.
In \S 3 we show the observed morphologies, and flux ratios in selected positions.
In \S 4 we present our LTE and non-LTE calculations, and perform comparisons with the observed morphologies and flux ratios.
In \S 5, we discuss the nature of shocked emission observed in these Herbig-Haro objects, and the applicability and limitations of three-color images.

\section{Data}

Archival data in four IRAC bands (3.6, 4.5, 5.8, and 8.0 \micron) were obtained for six Herbig-Haro objects: HH 211, HH 212, L 1448, L 1157, BHR 71 and HH 54. All the data had been reduced with the post-BCD pipelines developed by IPAC. The mean FWHM of the point response functions (PRFs) are 1.66, 1.72, 1.88 and 1.98'' for four bands, respectively.
Previous publications with these data include \citet{Tobin07} for L 1448; \citet{Looney07} for L1157; \citet{Walawender06} for HH 211; \citet{Neufeld09} for L1448, L1157 and BHR 71; and YL09 for HH 54. The median of the total integration time per pixel ranges 21--322 s among the objects. We did not stack all the post-BCD images available on archive, but their signal-to-noise are sufficient for our study, as shown in later sections.

Narrow-band images of H$_2$ 2.12 \micron ~ were obtained using HH 211, HH 212 and L 1448 using MOIRCS \citep{Ichikawa06, Suzuki08} on Subaru 8.2-m and ISAAC \citep{Moorwood98} on VLT 8.0-m. The MOIRCS observations of HH 211 and L 1448 were made on January 3, 2007.  The H$_2$ filter on MOIRCS provides a FHWM wavelength of 0.021 \micron.
The camera consists of two detectors with a FOV of 4'$\times$3' each with a pixel sampling of 0".117 pix.$^{-1}$, and since one of them is sufficient to cover our target, the other FOV was used for simultaneously obtaining the sky frames. Each object was observed with a Detector 1--2--2--1  sequence with an individual exposure of 180 s, providing a total on-source integration of 720 s. The standard data reduction (sky subtraction, flat fielding, field cosmetics, image stacking) was made using IRAF. Flux calibration was made using stars in the FOV and 2MASS catalog, with an accuracy of about 10 \%. The 1-$\sigma$ flux per pixel is 4.1 and 3.6$\times$10$^{-9}$ W m$^{-2}$ str$^{-1}$ for HH 211 and L 1448, respectively. The seeing of 0".5--0".6  is measured using stars in the same FOV.

The ISAAC images of HH 212 were obtained through ESO archive \citep{Takami06b}. These images are for two 2'.5 $\times$ 2'.5 FOV, covering an outflow length of about 4'. The exposure time of each frame and a median of total on-source integration time per pixel are 60 and 840 s, respectively. The FWHM wavelength of the filter is 0.028 \micron. Flux calibration was made using stars in the FOV and 2MASS catalog, with an accuracy of about 10 \%. The 1-$\sigma$ flux per pixel is 9$\times$10$^{-9}$ W m$^{-2}$ str$^{-1}$. The seeing of 0".4  is measured using stars in the same FOV.

\section{Results}
\subsection{Morphology}
\begin{figure*}
\epsscale{1.7}
\vspace{-0.5cm}
\plotone{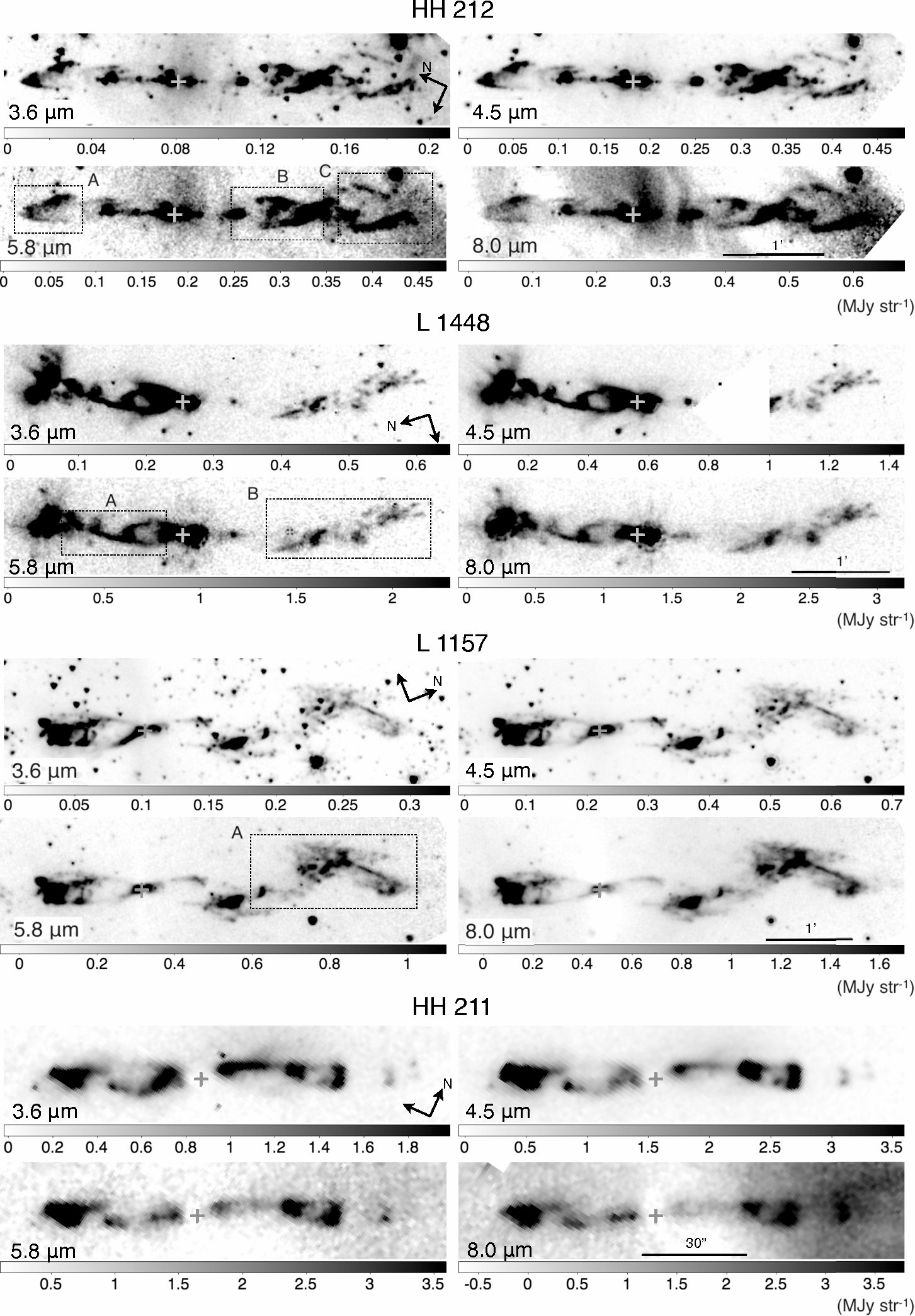}
\caption{IRAC Images of L1448, HH 212, L 1157, and HH 211. The flux is shown for each image with units of MJy str$^{-1}$. The cross indicates the position of the driving source \citep{LeeC08, LeeC09, Bachiller01, Girart01}.
Boxes with dashed lines indicate the regions shown in Figures \ref{fig_images3}--\ref{fig_images5}. Archival data is missing in the south part (0'.5--1') of the L1448 jet at 4.5 \micron. \label{fig_images1}}
\end{figure*}

Figure \ref{fig_images1} shows the image of the entire outflow in the four IRAC bands for L 1448, HH 212, L 1157 and HH 211. As shown in previous studies, these outflows show a collimated morphology with bow shocks, knotty, and bubble-like structures. The overall morphology in the four bands is similar, in particular 3.6 and 4.5 \micron. Even so, the morphology at short (3.6/4.5 \micron) and long (5.8/8.0 \micron) wavelengths are not identical as shown later in detail. L 1448 exhibits a conical nebulosity just north of the driving source, in particular at 3.6 and 4.5 \micron. As discussed by \citet{Tobin07}, it is explained as scattered continuum in the outflow cavity. A similar nebulosity close to the driving source is also visible on both sides of L 1157 and HH 211. \citet{Tobin07} concluded that the nebulosity at the north end of the  L 1448 outflow is also due to scattered continuum, by a different YSO located there.

\begin{figure*}
\epsscale{1.7}
\vspace{-0.5cm}
\plotone{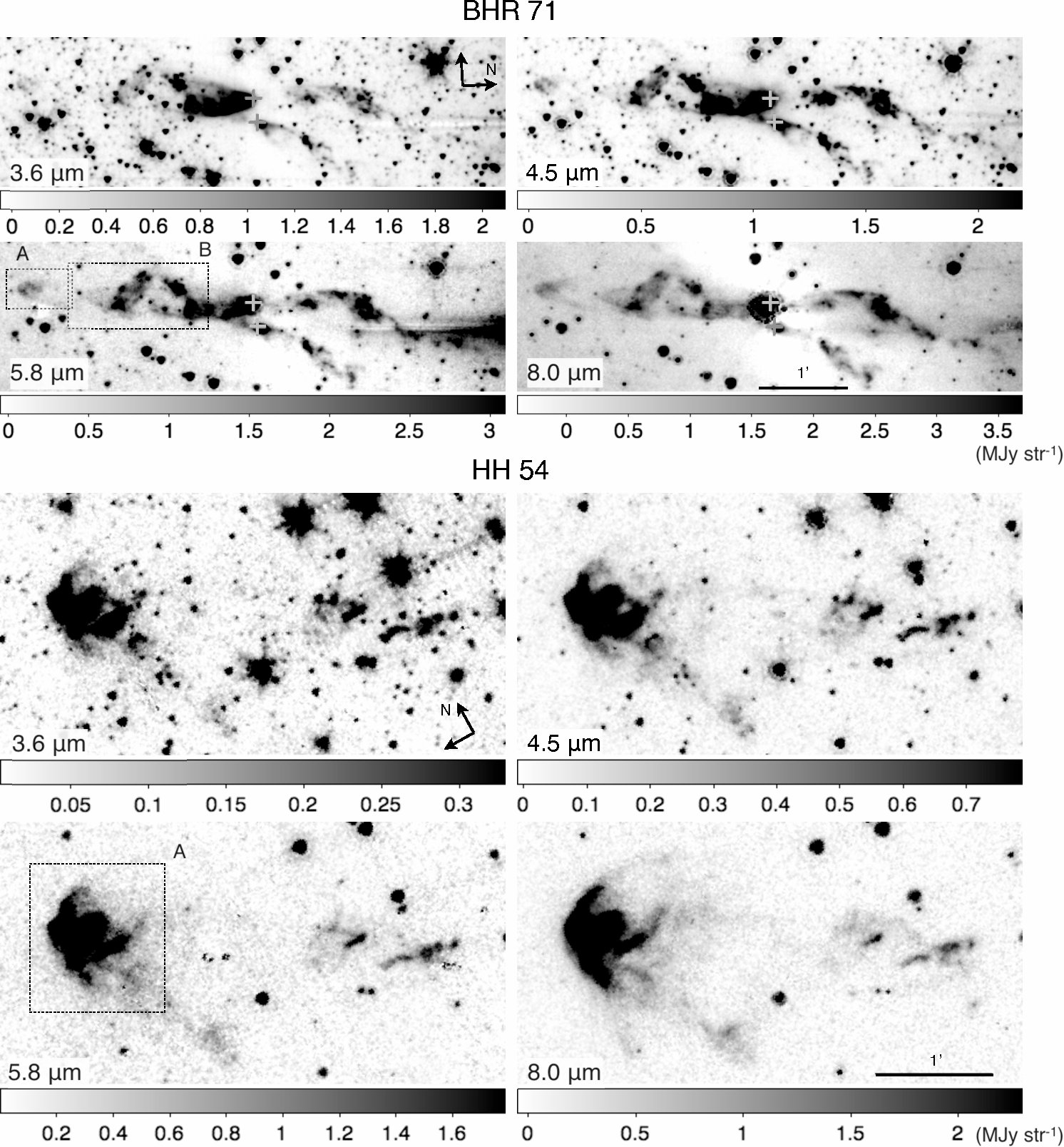}
\caption{Same as Figure \ref{fig_images1} but for BHR 71 and HH 54. The position of the driving sources of BHR 71 are based on \citep{Chen08}. The driving source of the HH 54 outflow is not clear \citep[see][]{Ybarra09}. \label{fig_images2}}
\end{figure*}

Figure \ref{fig_images2} shows the images of BHR 71 and HH 54. The BHR 71 outflow shows a collimated morphology like the other outflows in Figure \ref{fig_images1}, while the HH 54 outflow exhibits a bright, large ($\sim$1') bow shock in the northwest and a faint collimated flow toward the southwest. The head of the HH 54 bow shock consists of several knotty structures, also shown in the 4.5-\micron~ image in YL09. In contrast to the other objects shown in Figure \ref{fig_images1}, the overall morphology of BHR 71 and HH 54 is different at short (3.6--4.5 \micron) and long wavelengths (5.8--8.0 \micron). In BHR 71, the short wavelengths show a conical nebulosity close to the driving source more prominently, while the longer wavelengths show the outflow emission in more extended regions. Again, the former is presumably due to scattered continuum in the outflow cavity. In HH 54, the emission at shorter wavelengths  is enhanced at the head of the bow shock, while emission at the wakes is also seen at longer wavelengths.

\begin{figure*}
\epsscale{1.6}
\vspace{-0.5cm}
\plotone{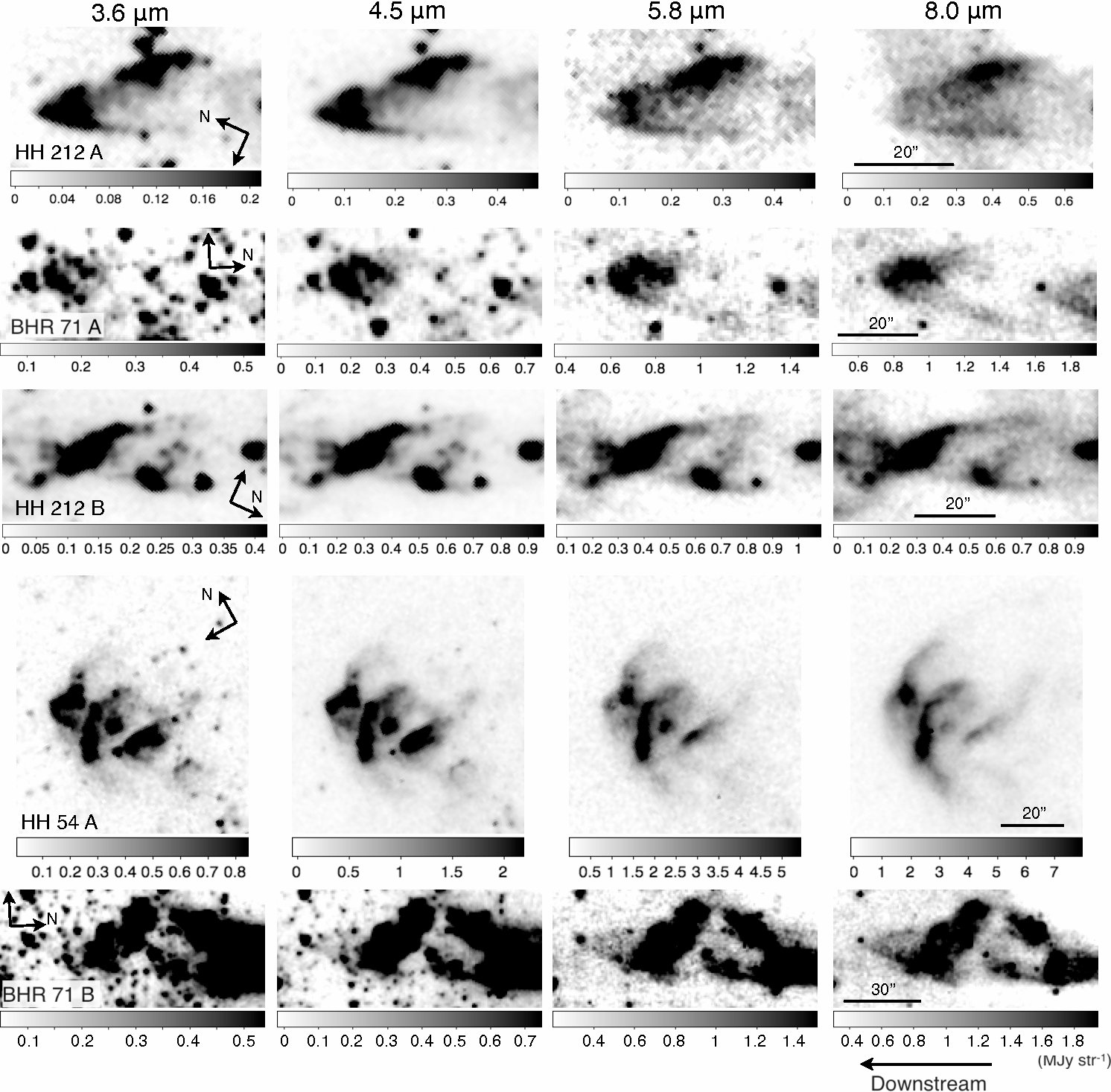}
\caption{Bow shocks and similar shock structures observed in HH 212, BHR 71 and HH 54. All the figures are rotated to align the direction of the down stream to the left. \label{fig_images3}}
\end{figure*}
\begin{figure*}
\epsscale{1.6}
\vspace{-0.5cm}
\plotone{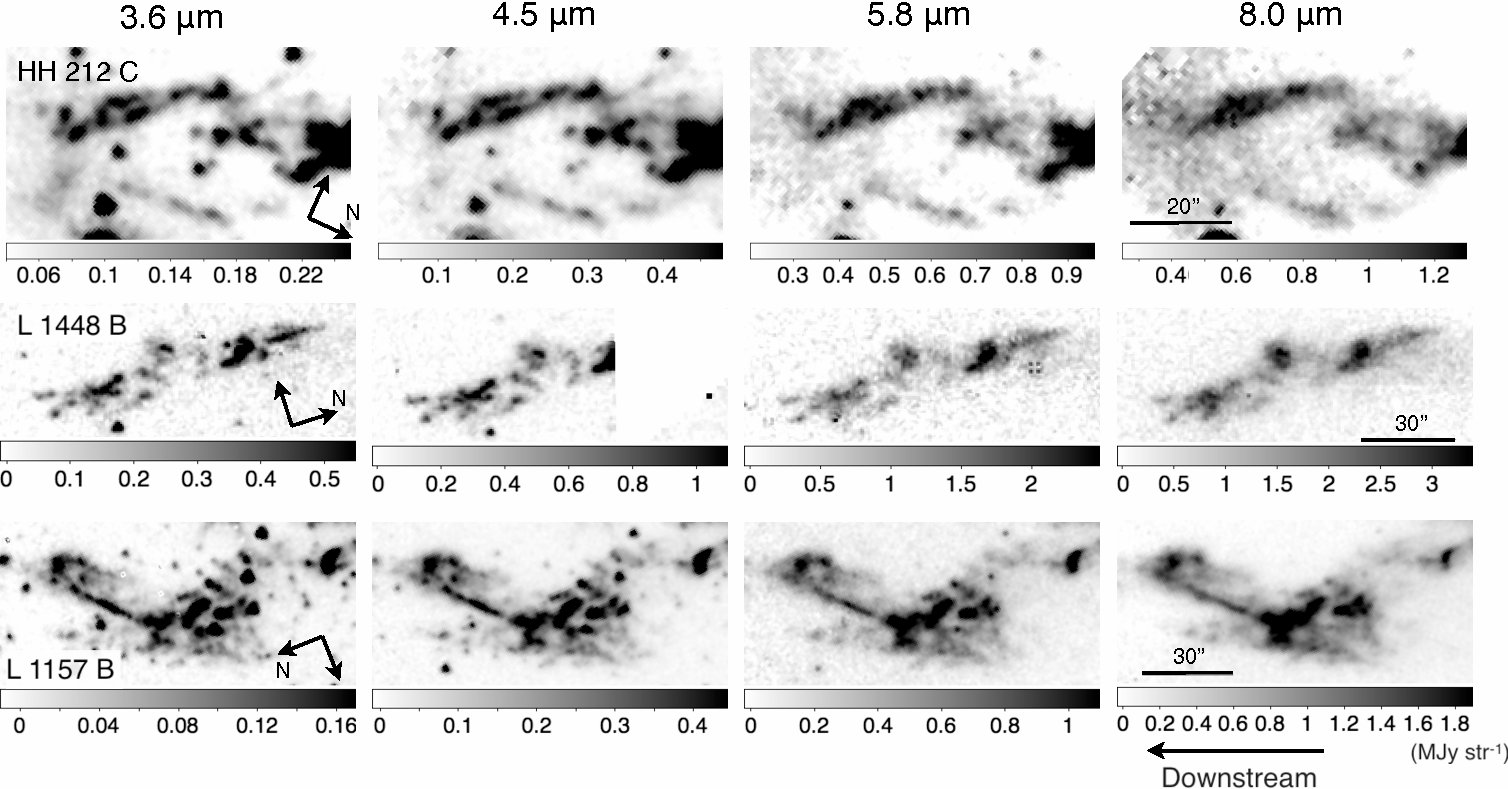}
\caption{Clumpy shock structures at selected regions in HH 212, L 1448 and L 1157 B. All the figures are rotated to align the direction of the down stream to the left. Archival data is missing near the right end of the L1448 B at 4.5 \micron. \label{fig_images4}}
\end{figure*}

To show the different morphologies in the four bands more clearly, we select several regions and show close-up views in Figures \ref{fig_images3} and \ref{fig_images4}. Figure \ref{fig_images3} shows morphologies of bow shocks at four IRAC bands observed in HH 212, BHR 71, and HH 54.  In HH 212 A, the emission at shorter wavelengths is enhanced at the head of the bow shocks, and also in a  bright wake in the west part of the bow shock. In contrast, emission at longer wavelengths is enhanced at both wakes of the bow shock. The emission at the head of the bow shock is absent at 8.0 \micron. In HH 212 B, BHR 71 A, and HH 54 A, the images at all four bands exhibit bright peaks at the head of the bow shocks, and the 5.8 and 8.0 \micron~ images also show faint extended emission toward the wakes. In BHR 71 B, the bow shocked emission is clearly seen only at 5.8 and 8.0 \micron, and it is marginal at the other wavelengths. In all the regions described above, the wakes of the bow shocks are most prominent at 8.0 \micron. These tendencies are shown in shocked emission with hydrodynamical models by \citet{Smith05}.
The morphology at 3.6 and 4.5 \micron~ is similar in all of these structures.

Figure \ref{fig_images4} shows the knotty structures seen in HH 212, L 1448, and L 1157. Emission at 3.6 and 4.5 \micron~ exhibits patchy structure in these shocks more clearly, while it is more blurred at longer wavelengths (5.8--8.0 \micron), exhibiting diffuser emission components between emission peaks. The observed morphologies at 3.6 and 4.5 \micron~ are almost identical.
HH 212 C shows a bow-shock morphology, but the different contrast between the head and wakes is not clearly seen, perhaps due to the fact that emission is associated with only some parts of the wakes.

\begin{figure*}
\epsscale{1.6}
\vspace{-0.5cm}
\plotone{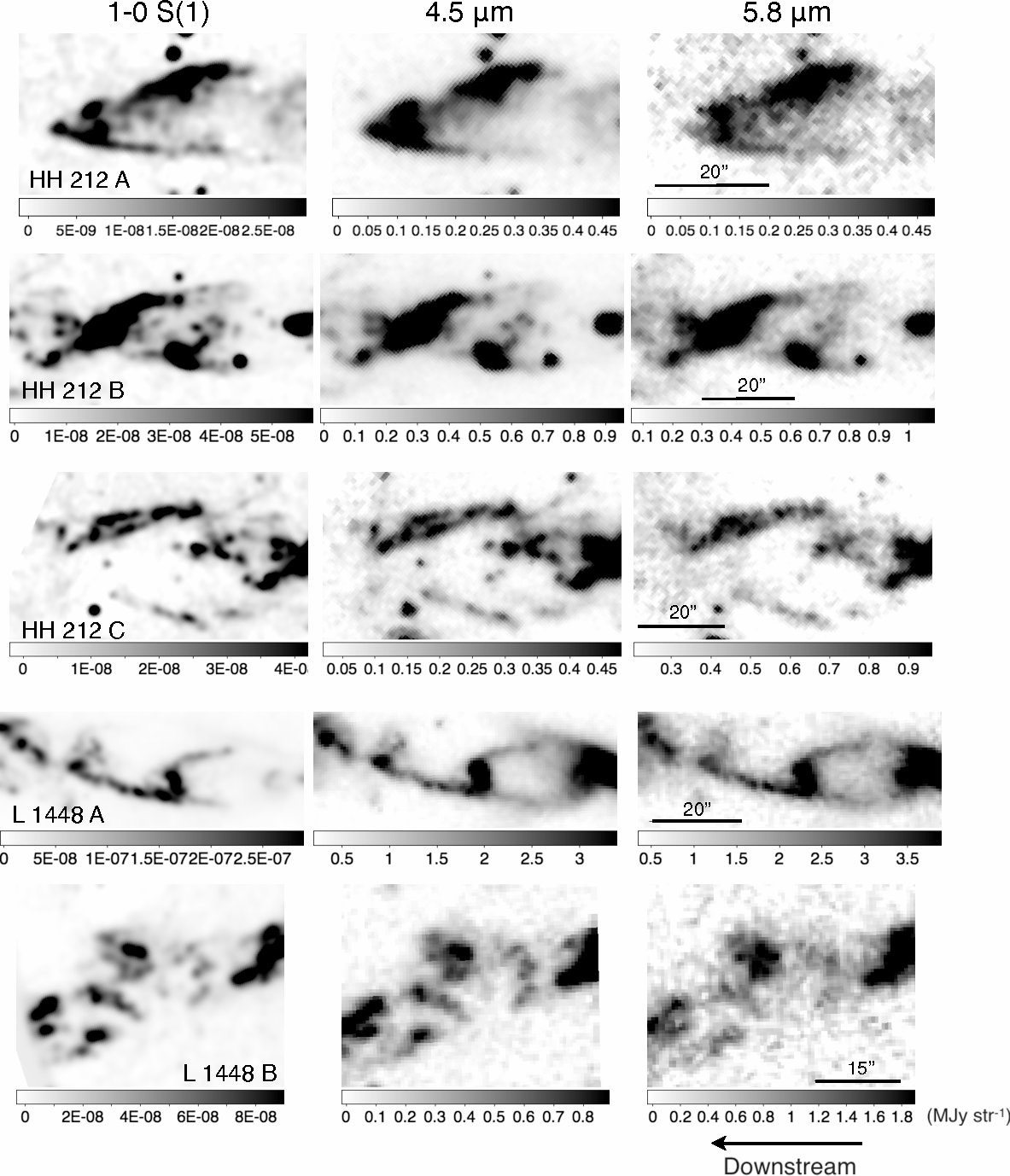}
\caption{1-0 S(1) 2.12\micron, IRAC 4.5 and 5.8 \micron~ images of selected regions in HH 212 and L 1448. The flux units are W m$^{-2}$ str$^{-1}$ for H$_2$, and MJy str$^{-1}$ for IRAC images. All the figures are rotated to align the direction of the down stream to the left. \label{fig_images5}}
\end{figure*}

Figure \ref{fig_images5} shows images of 1-0 S(1), together with the IRAC 4.5 and 5.8 \micron~ emission at the selected regions in HH 212 and L 1448. To compare the 1-0 S(1) morphology with the others, we convolve the 1-0 S(1) images with a gaussian to match the angular resolution to IRAC 4.5 \micron. These show that the morphology at 1-0 S(1)  is similar to the 4.5 \micron~ emission rather than 5.8 \micron. Together with Figures \ref{fig_images3} and \ref{fig_images4}, we conclude that the mophologies in 1-0 S(1), IRAC 3.6 \micron~ and 4.5 \micron~ are similar to each other.


\subsection{Flux Ratios}
\begin{figure*}
\epsscale{2}
\vspace{-0.5cm}
\plotone{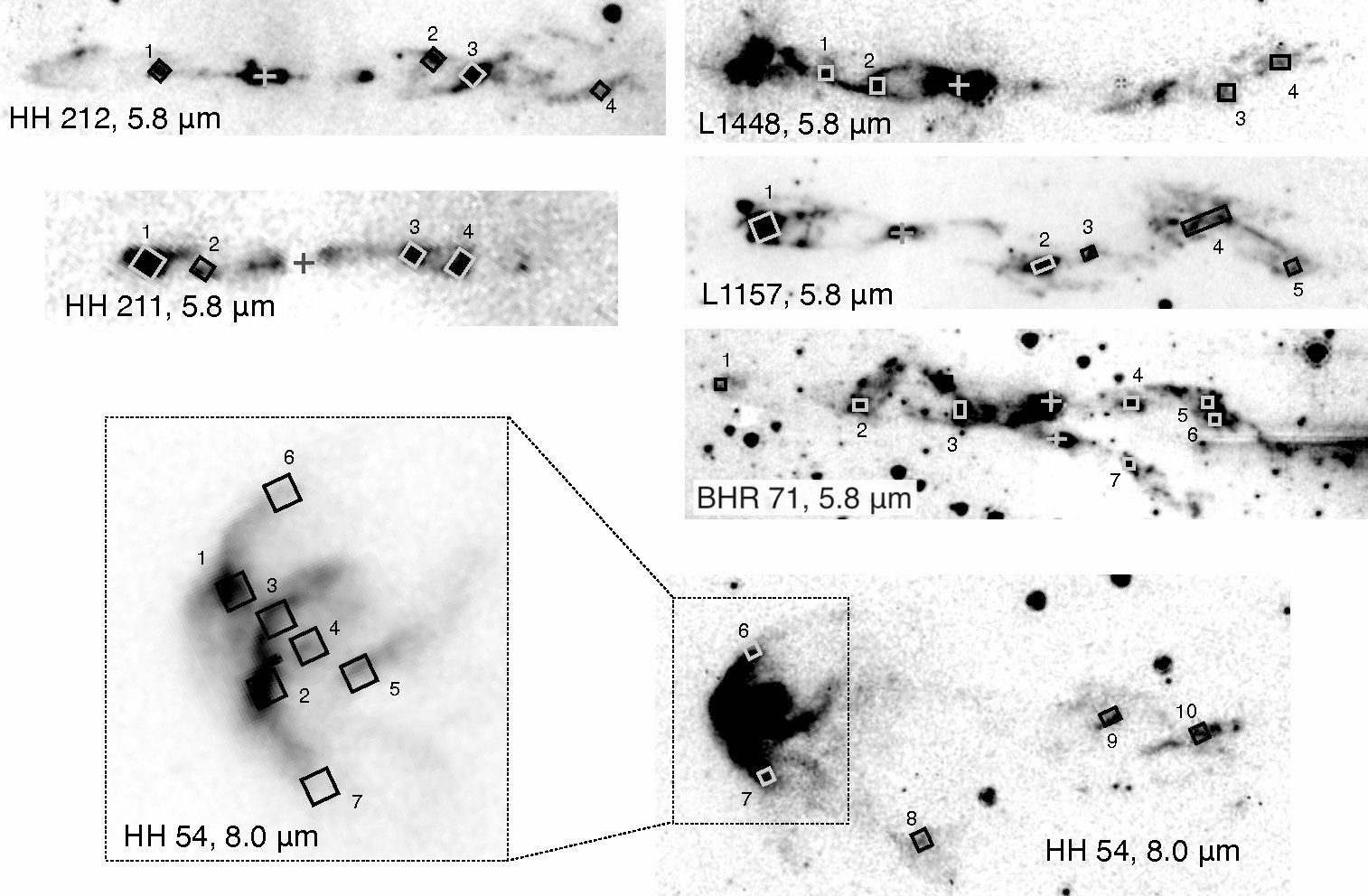}
\caption{Positions where we measure spectral energy distributions in IRAC bands. See text for details. \label{fig_images6}}
\end{figure*}

To perform comparisons with model calculations in detail, we measured the fluxes at the positions shown in Figure \ref{fig_images6}. 
These measurements were made carefully with the following two caveats. First, shocked emission in some objects is severely contaminated from diffuse emission in particular at 8.0 \micron, presumably due to PAH \citep[see, e.g.,][]{Reach06}. Some shocked emission seen in at 3.6 and 4.5 \micron~ is contaminated by emission from stars, and by scattered continuum near the target protostar. We avoided such regions to allow reliable discussion in later sections. For each point shown in Figure \ref{fig_images6}, we also carefully subtracted the diffuse background emission by measuring it at x- and y-directions and fitting them with a 1-D plane.

Secondly, the different PRFs between the four bands would cause systematic errors on flux ratios. The IRAC Data Handbook issued by Spitzer Science Center suggests to perform convolution of images to cancel out this effect. We thus convolve the 3.6 and 4.5 \micron~ images with PRFs at 5.8 and 8.0 \micron; 5.8 \micron~ images with PRFs at 4.5 and 8.0 \micron; and 8.0 \micron~ images with PRFs at 4.5 and 5.8 \micron. This process yields an effective angular resolution of $\sim$4" for each image. Note that the PRFs at 3.6 and 4.5 \micron~ are almost identical, thus we do not correct the difference between these two PRFs.

\begin{figure*}
\epsscale{2}
\vspace{-0.5cm}
\plotone{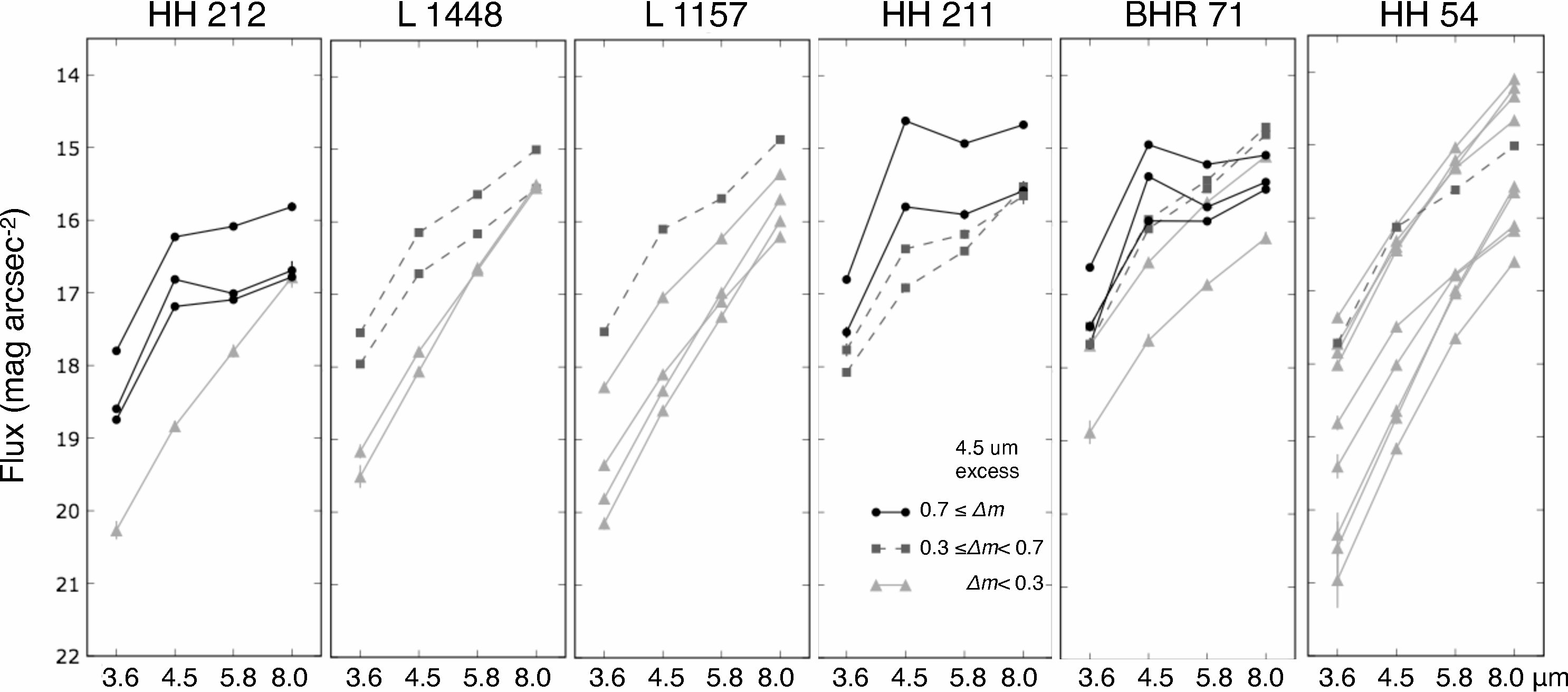}
\caption{Flux observed in Figure \ref{fig_images6}. Different grayscale and dots are used for different degree of excess emission at 4.5 \micron. The excess $\Delta m_{4.5}$ is defined as ($m_{3.6}$+$m_{5.8}$)$\times$0.5--$m_{4.5}$. The error bars are shown only for those larger than the size of the dots. \label{fig_obs_SEDs}}
\end{figure*}

Table \ref{tbl_obs_IRAC} and  Figures \ref{fig_obs_SEDs} show the measured flux density at individual positions. Figure \ref{fig_obs_SEDs} shows that the observed spectral energy distributions (SEDs) fall into two categories: (A) those in which the $magnitude$ monotonically decrease as the wavelength increases; and (B) those with excess emission at 4.5 \micron. We define the 4.5 \micron~ excess as $\Delta m_{4.5} \equiv (m_{3.6}+m_{5.8}) \times 0.5 - m_{4.5} $, and use different grayscales in Figures \ref{fig_obs_SEDs}  for different magnitude of excess. The figure shows that the type-B SEDs tend to be brighter than type-A SEDs in each.
The color gradient of type-A SEDs ranges between $m_{3.6}-m_{8.0}$=2.5--5.5, and this corresponds the 3.6/8.0-\micron~ flux ratio of 0.03--0.4 based on units of MJy str$^{-1}$. Thus, the $flux$ monotonically increases with wavelength in type-A SEDs. Those in L 1157 and HH 54 show that the SEDs in shallow color gradients tend to be brighter than the steeper SEDs. 

\begin{figure*}
\epsscale{1.4}
\vspace{-0.5cm}
\plotone{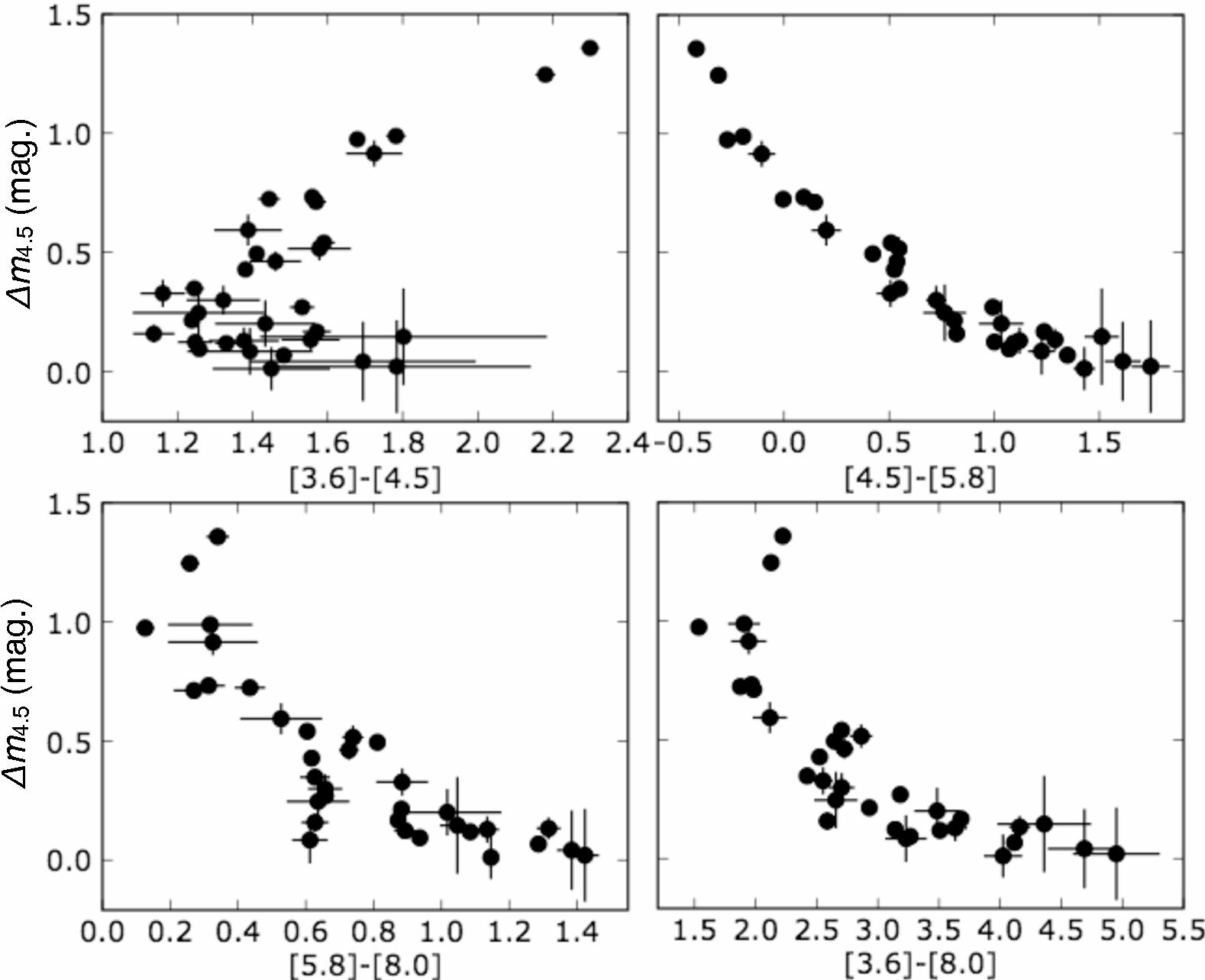}
\caption{Correlation between the observed IR color and excess at 4.5 $\micron$.  
The error bars are shown only for those larger than the size of the dots.
  \label{fig_obs_excess_corr}}
\end{figure*}

Figure \ref{fig_obs_excess_corr} shows correlations between the observed IRAC colors and the 4.5-\micron~ excess. Any color shows a good correlation with the 4.5-\micron~ excess, implying that the entire SED is characterized by this excess. This issue is discussed in detail in \S 5.3.

\begin{figure*}
\epsscale{1.4}
\vspace{-0.5cm}
\plotone{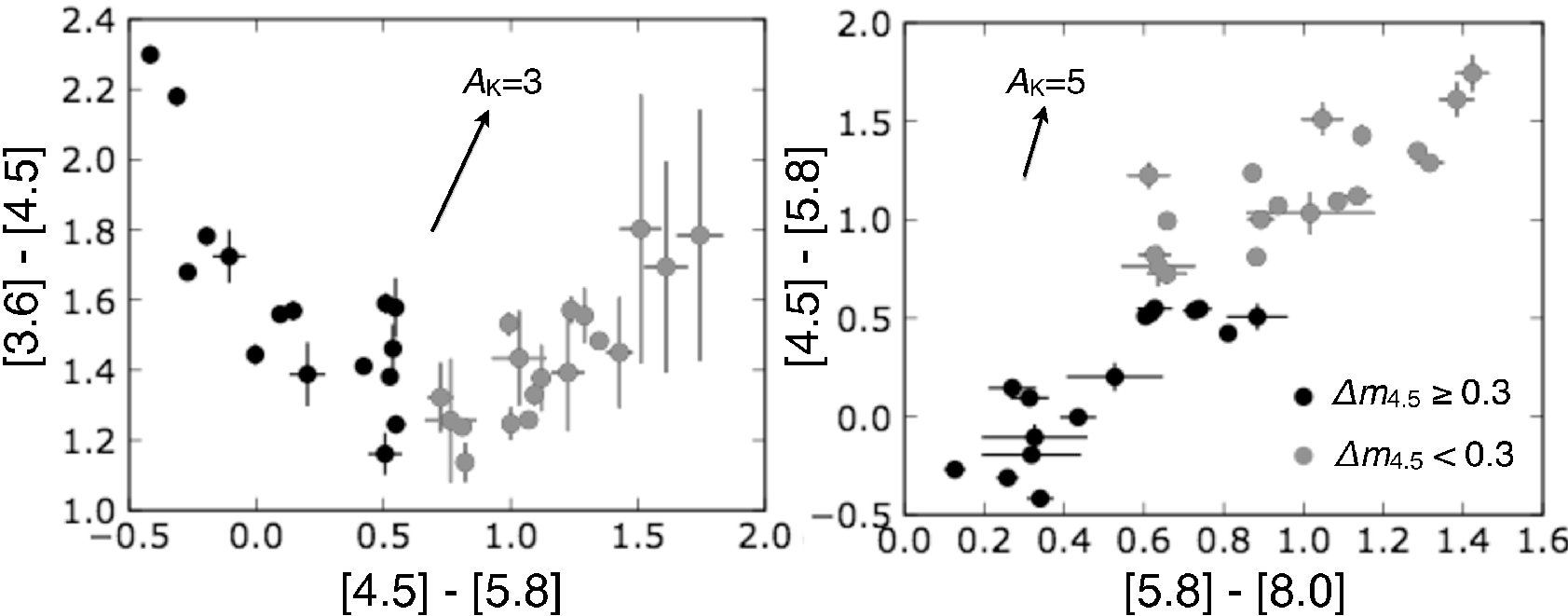}
\caption{Color-color diagrams of the observed flux. The black and gray dots indicate the results with excess 4.5-\micron~ magnitude $\Delta m_{4.5} \ge 0.3$ and $<0.3$, respectively. The arrows for extinction is based on Chapman et al. (2009) for $A_K > 2$.  \label{fig_cc_vs_ext}}
\end{figure*}

Figure \ref{fig_cc_vs_ext} show color-color diagrams of these SEDs. As expected from Figure \ref{fig_obs_excess_corr}, good correlations are observed between [3.6]--[4.5], [4.5]--[5.8], and [5.8]--[8.0] colors. The figure also shows the colors of extinction measured in molecular clouds by \citet{Chapman09}. Although different extinction would produce a color gradient similar to type-A SEDs, the slope of the color
of type-A SEDs in the color-color diagrams in Figure \ref{fig_cc_vs_ext} is shallower than that expected from extinction. Furthermore, the correlation between [4.5]--[5.8] and [5.8]--[8.0] can be fitted with a single line with both type-A and type-B SEDs. These suggest that the observe color gradient is due to different excitation conditions in shocks, not extinction.

\begin{figure*}
\epsscale{1.4}
\vspace{-0.5cm}
\plotone{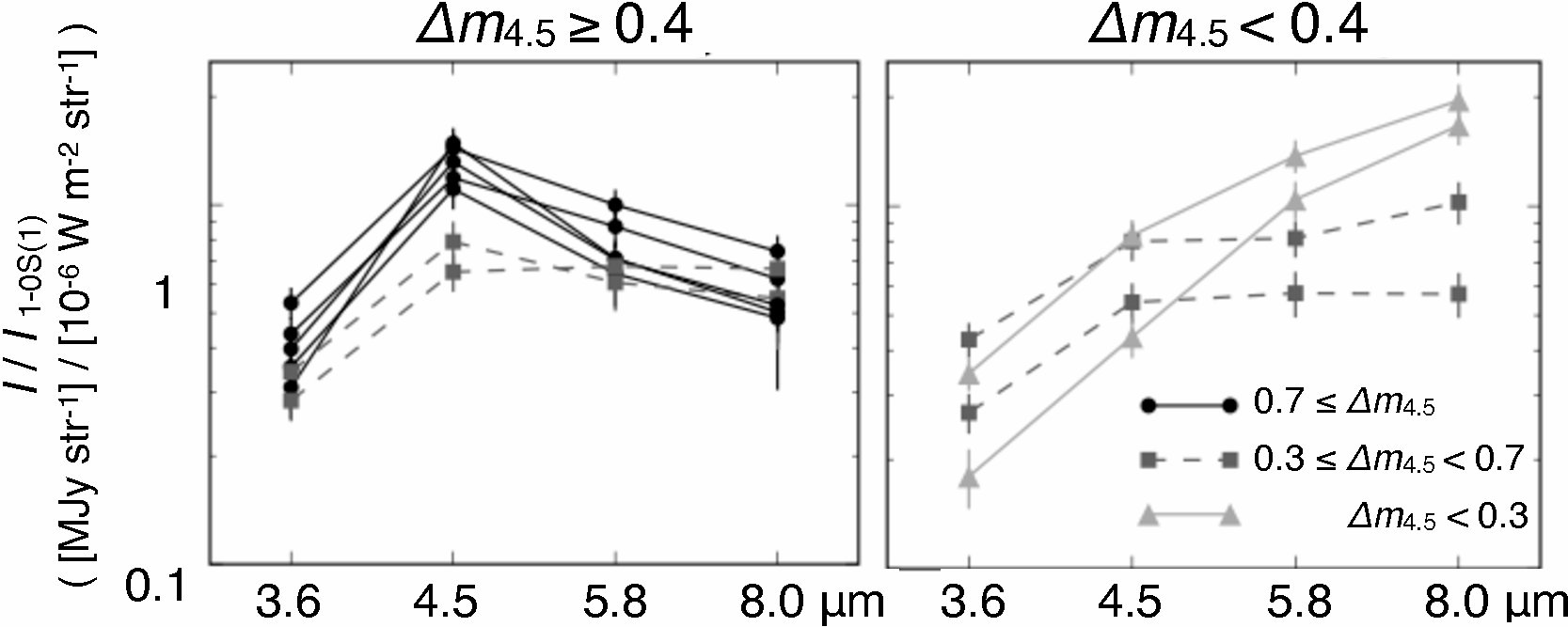}
\caption{Observed IRAC flux normalized by the 1-0 S(1) flux. The left and right figures show SEDs with different excess 4.5-\micron~ magnitudes ($\Delta m_{4.5} \ge 0.4$ and $<0.4$, respectively). Different grayscale and dots are used for different degree of excess emission at 4.5 \micron. \label{fig_obs_SEDs_per10S1}}
\end{figure*}

In addition to the IRAC fluxes, we also measured the 1-0 S(1) flux at positions where the data are available. Before the flux measuremets, these are convolved with IRAC PRFs of 4.5, 5.8 and 8.0 \micron~ to approximately match the angular resolutions. The results are shown in Table \ref{tbl_obs_10S1}. Figure \ref{fig_obs_SEDs_per10S1} shows the IRAC SEDs at these positions normalized to 1-0 S(1) emission. The 4.5 \micron/1-0 S(1) flux ratio is higher  for SEDs with a larger 4.5-\micron~ excess $\Delta m_{4.5}$. More quantitatively, 
the flux ratio in SEDs with $\Delta m_{4.5} > 0.3$ is larger than those with $\Delta m_{4.5} < 0.3$, typically by a factor of $\sim$2.

\section{Comparisons with Modeled H$_2$ Emission}

We compare the results in \S 3 with calculations of thermal H$_2$ emission. These calculations include:- (1) LTE calculations with 245 energy levels, including 207 lines in the four IRAC bands (\S 4.1); and (2) non-LTE calculations with 36--49 levels, including 32--45 lines in the four IRAC bands (\S 4.2). The non-LTE calculations are made for collisional particles for two extreme cases: H+He and H$_2$+He. For the temperature structure, we use simplified shock models with a power-law cooling rate ($\Lambda \propto T^s$), in addition to isothermal cases.

The comparison between observations are made with SEDs and color-color diagrams. Although the flux ratio with 1-0 S(1) could be useful for testing models  (Figure \ref{fig_obs_SEDs_per10S1}), we do not perform such comparisons due to uncertainties in extinction, and the absolute flux calibration of the IRAC data for extended sources (see IRAC Data Handbook).

\subsection{LTE models}
\subsubsection{Description and basic characteristics}
The list of energy levels and transitions are obtained from models of photo-dissociation regions by \citet{DB96}. 
This covers 245 energy levels at $E/k$=0--43000 K, and among them, we use all of the transitions at  3.09--10.2 \micron~ (207 lines). For the  $A$-coefficients we use recent calculations by \citet{Wolniewicz98}. For isothermal cases, the IRAC flux is calculated from each spectrum with the following equation:

\begin{equation}
I_{IRAC} =  c_{I} \sum_{j,k} R_{I} (\Lambda_{jk}) ~A_{jk}   \frac{n_{H_2} A \Omega~ g_j e^{-\frac{E_j}{kT}}}{ Z(T)},
\end{equation}
where $I_{IRAC}$ is the flux observed in a IRAC band; $C_I$ is the conversion factor from electrons to MJy str$^{-1}$ (see IRAC Data Handbook); $j$ and $k$ are the upper and lower levels of transitions, respectively; $R_I (\Lambda_{jk})$ is the spectral response function at the wavelength $\Lambda_{jk}$; $A_{jk}$ is the Einstein A-coefficient; $n_{H_2}$ is the column density of molecular hydrogen; $A$ is the area of the telescope mirror (0.4636 m$^2$); $\Omega$ is the angular area corresponding to a pixel (1".22$\times$1".22 or 3.498$\times 10^{-11}$ str); $g_j$ is the statistical weight of the level $j$; $E_j$ is the upper level energy; and $Z$($T$) is the partition function.

In addition to the isothermal cases, we also calculate the IRAC fluxes assuming the temperature structures used in \citet[][hereafter B88]{Brand88}, and \citet[][hereafter NY08]{Neufeld08}. B88 developed a simplified dissociative $J$-shock model with a power-law cooling rate $\Lambda = \Lambda_0 T^s$ to explain infrared H$_2$ emission at a broad range of wavelengths. The column density at the energy level $j$ is given as:
\begin{eqnarray}
N_j/g_j &\propto&  \int e^{-T_j/T} ~[Z(T)\Lambda(T)]^{-1}~ dT,  \nonumber \\
              &\propto&  \int e^{-T_j/T} ~Z(T)^{-1}~ T^{-s}~dT, 
\end{eqnarray}
where $T_j = E_j/k$.
The authors provide an approximate formula for the population of each energy level of:-
\begin{eqnarray}
N_j/g_j &\propto&  T_j^{-s} - (T_j + T_v)^{-s}, 
\end{eqnarray}
with  $T_v$=6000 K. The B88 model has been successful in roughly explaining near-infrared spectra of  H$_2$ emission in shocks \citep[e.g.,][]{Brand88, Everett95, Richter95, Takami06a}. The power index $s$ is often close to 4.7,  i.e., cooling dominated by molecular hydrogen \citep{Brand88}. This model can also explain the relatively constant temperature measured at a variety of shock regions with near-infrared lines \citep[$T \sim 2000$ K, e.g.,][]{Burton89, Eisloffel00, Nisini02, Takami06a, Beck08}. The NY08 model is essentially the same as B88, but Equation (2) is numerically integrated with the upper limit of temperature $T$=4000 K. This temperature approximately correspond to the dissociation temperature of H$_2$ molecules \citep[e.g.,][]{Lepp83}. This model has been successfully applied to H$_2$ spectra at the 5--28 \micron~ range in some shocks, including the supernova remnant IC 443 (NY08) and several Herbig-Haro objects \citep{Neufeld09}. We set the lowest temperature of 30 K for numerical integration.

\begin{figure*}
\epsscale{2}
\plotone{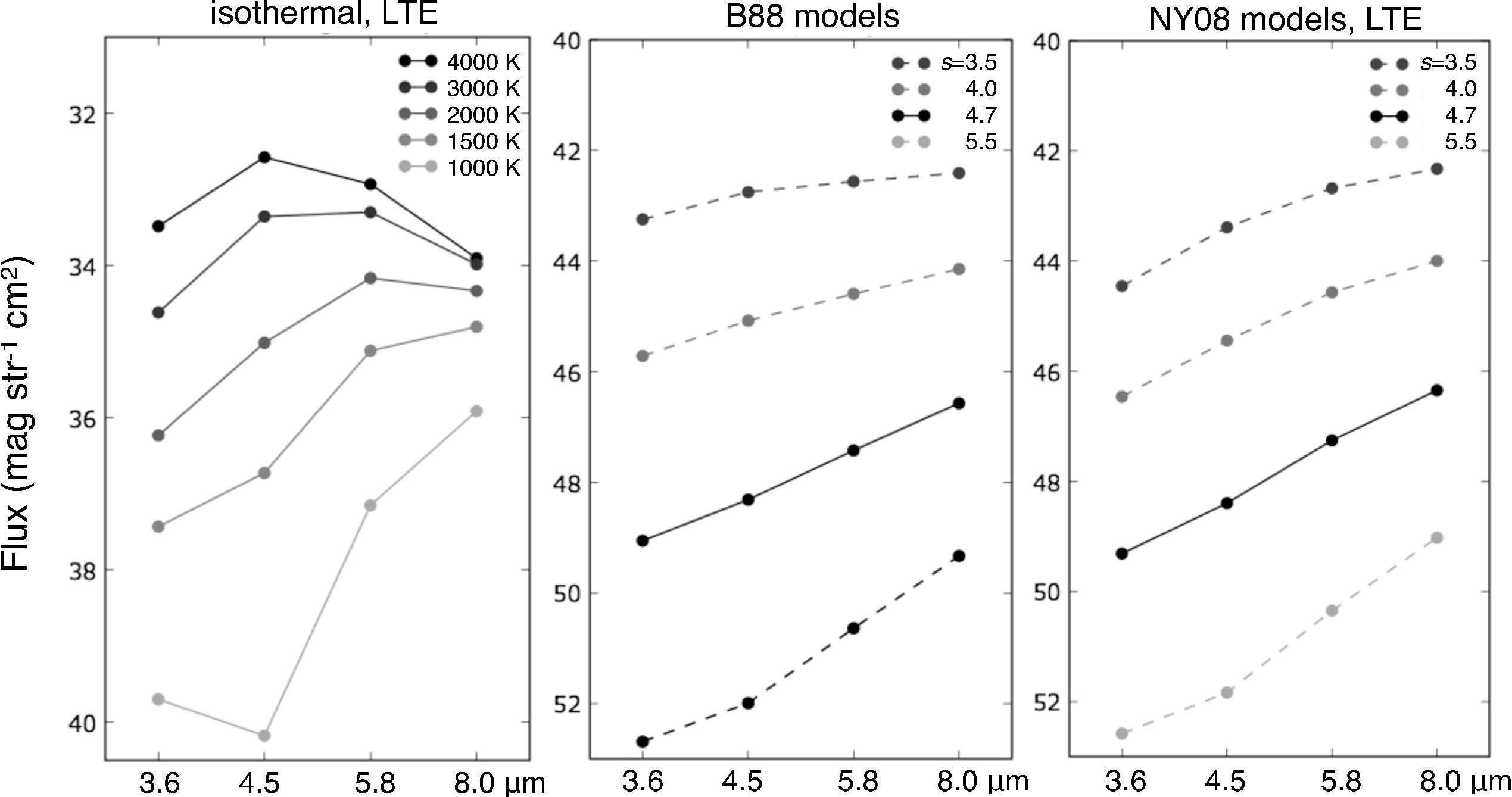}
\caption{Spectral energy distributions for LTE models per unit column density of H$_2$. (left) For isothermal cases at $T$=1000--4000 K.  (middle) For the Brand et al. (1988) models with different power indexes of the cooling function. Normalized by the column density of $excited$ H$_2$, since the models yields an infinite column density for the ground state (see Equation 3). (right) Same as the middle but for the Neufeld \& Yuan (2008) models. \label{fig_LTE_SEDs}}
\end{figure*}

Figure \ref{fig_LTE_SEDs} shows the SEDs of LTE H$_2$ for isothermal cases, and B88 and NY08 models with different power indexes for the cooling function. For B88 and NY08, the flux monotonically changes with wavelength, exhibiting a tendency similar to the type-A SEDs described in \S 3.2. The slope of the SEDs is steeper with a larger power index $s$ because of the contribution of gas at low temperatures to longer wavelengths. The B88 and NY08 models provide almost the same results at large $s$, while the SED is steeper for NY08 with small $s$, due to a smaller contribution of high-temperature slabs to the latter. None of the LTE models discussed here explains the type-B SEDs, which show excess emission at 4.5 \micron.

\begin{figure}
\epsscale{0.8}
\plotone{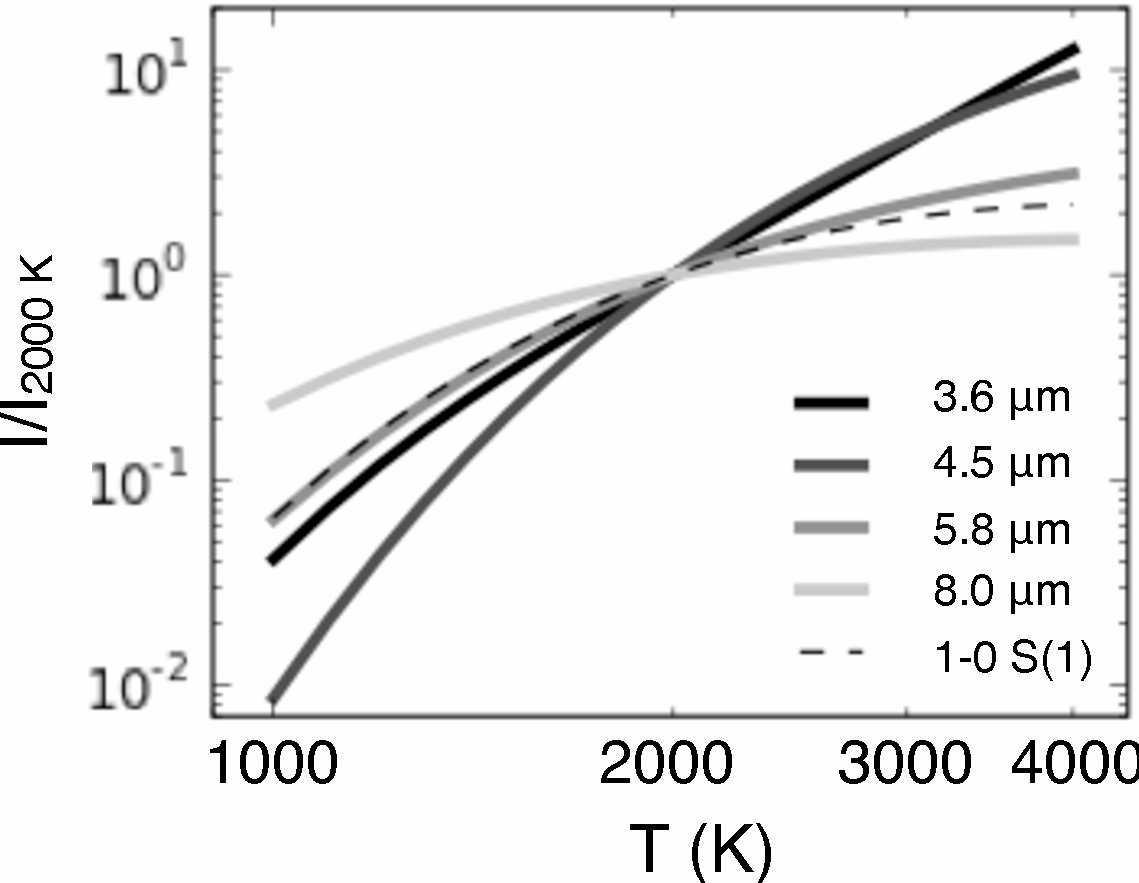}
\caption{IRAC and 1-0 S(1) fluxes for LTE-H$_2$ as a function of temperature. The flux for each band or line is normalized to that at $T$=2000 K. \label{fig_LTE_curve}}
\end{figure}

Figure \ref{fig_LTE_curve} shows the IRAC and 1-0 S(1) fluxes as a function of temperature. These are normalized at 2000 K, i.e., the typical temperature observed in shocks based on ground-based infrared spectroscopy \citep[e.g.,][]{Burton89, Eisloffel00, Nisini02, Takami06a, Beck08}. The figure shows that, compared with 1-0 S(1), the 3.6 and 4.5 \micron~ emission enhances regions at higher temperatures; the 8-\micron~ emission does regions at lower temperatures; and the 5.8-\micron~ emission traces regions similar to 1-0 S(1). This contrasts with the observed tendency for the 3.6 and 4.5 \micron~ emission to show morphologies similar to 1-0 S(1), rather than 5.8-\micron~ (\S 3.1).

\subsubsection{Color-color diagrams}

\begin{figure*}
\epsscale{2}
\plotone{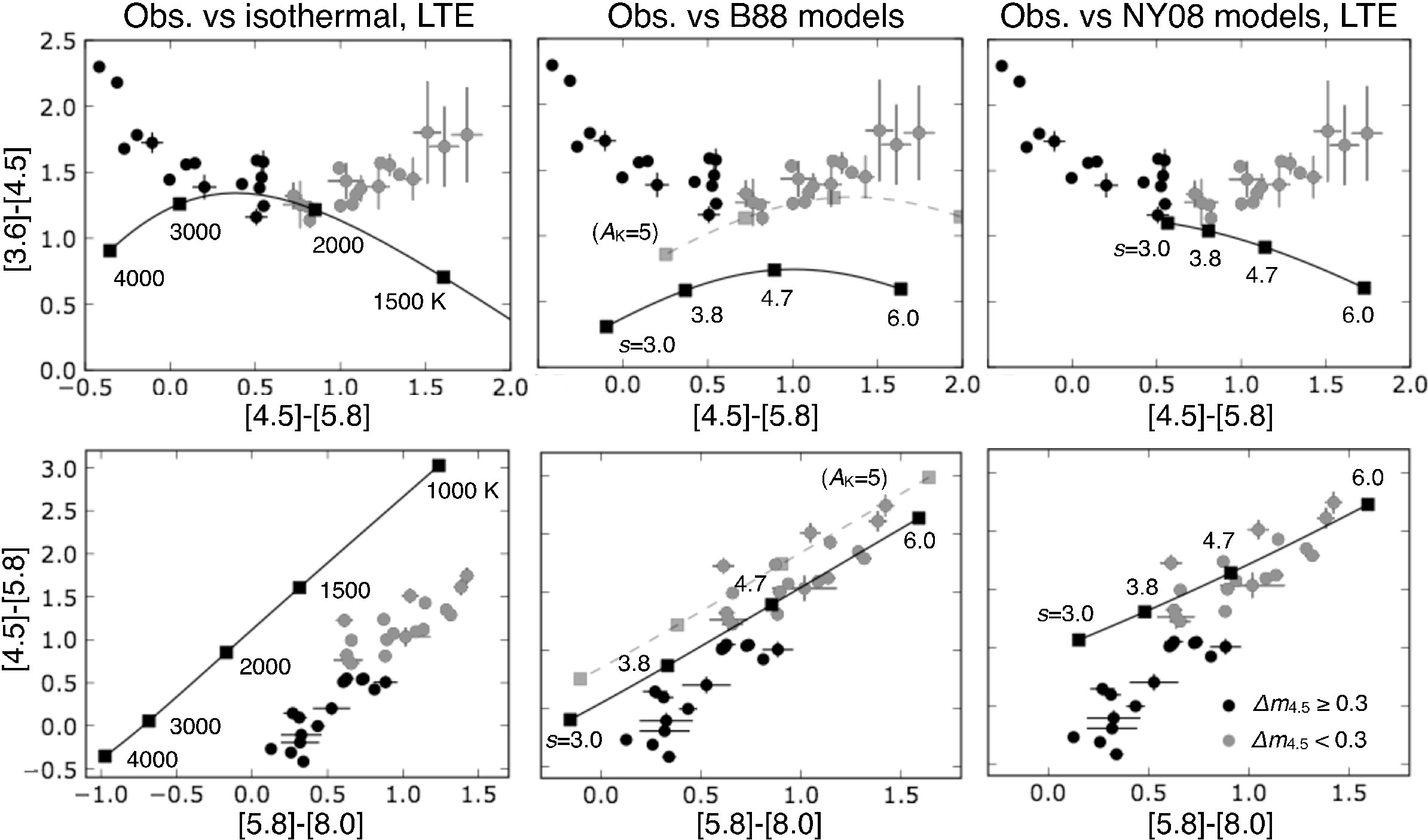}
\caption{Same as  Figure \ref{fig_cc_vs_ext} but the colors of LTE models are overplotted with a solid curve/line and square dots.  (left) At a single temperature.  (middle) For the Brand et al. (1988) models with different power indexes of the cooling function. The gray dashed line with square dots indicate the case for $A_K$=5.  (right) For Neufeld \& Yuan (2008) models with different power indexes of the cooling function. 
\label{fig_cc_LTE}}
\end{figure*}

Figure \ref{fig_cc_LTE} shows color-color diagrams of observed and modeled fluxes. 
In the [4.5]--[5.8] vs. [3.6]--[4.5] diagrams, the models could explain the observed colors for only a fraction of SEDs ([4.5]--[5.8]=0.4--1.1).
Isothermal models at 2000--3000 K and the NY08 models with $s \sim 3.0$ explain the observed flux at [4.5]--[5.8]$\sim$0.5.  The B88 models explain the observed flux ratio at [4.5]--[5.8]$\sim$1.0 if we assume a large extinction ($A_K$=5). In any case, the model curves cannot entirely fit the distribution of observed colors in the diagrams. In particular, the observed fluxes show [3.6]--[4.5]$\gtrsim$1.3, while any LTE models flux show [3.6]--[4.5]$\lesssim$1.3 if extinction is not included.

Regarding the  [5.8]--[8.0] vs. [4.5]--[5.8] diagrams, the B88 and NY08 models explain the color of the observed SEDs with little or no 4.5 \micron~ excess ($\Delta m_{4.5} < 0.3$) well. The SEDs at a larger 4.5 \micron~ excess show smaller [4.5]--[5.8] colors by up to $\sim$1.0 mag. The isothermal models cannot explain either type-A or type-B SEDs.

\subsection{Non-LTE models}
\subsubsection{Description and Characteristics}
Following YL09,
we adopt  $A$-coeffcients provided by \citet{Wolniewicz98}, and collisional rate coefficients with H$_2$ and He by \citet{LeBourlot99}. For collisional rate coefficients with H, we adopt \citet{Wrathmall07} and \citet{LeBourlot99}, and show the results separately.
According to \citet{Wrathmall07}, they provide rate coefficients with a better accuracy than \citet{LeBourlot99} because of improved representation of the vibration eigen functions. In contrast, coefficients provided by \citet{LeBourlot99} can explain the observed SEDs better as shown in \S 4.2.2.
Table \ref{tbl_L99W07_summary} summarizes the 
energy levels included in the online tables of \citet{Wrathmall07} and \citet{LeBourlot99}. Considering the completeness of the transitions covered for ortho- and para-H$_2$, we include the lowest 49 levels ($E/k=2 \times 10^4$) in case of H+He collisions, and 36 levels ($E/k=1.69 \times 10^4$ K) in case of H$_2$+He collisions. 
For the He/H abundance ratio, we adopt 0.098 based on \citet{Dappen00}.

Transitions and rate coefficients between ortho- and para-H$_2$ are not included in the above physical parameters, hence we calculate the populations of ortho- and para-H$_2$ separately, and combine them assuming an ortho/para ratio of 3. This ratio is corroborated by near-infrared spectroscopy of shocked regions \citep[see, e.g.,][]{Everett95, Giannini01, Nisini02, Takami06a, Beck08}, and explained by chemical processes of grain surfaces sufficient for thermal equilibrium \citep{LeBourlot99}. 

The calculations are made for two cases: for isothermal and the NY08 temperature structure (\S 4.1.1). Due to the limited number of transitions included, non-LTE calculations produce large errors at large temperatures (or small $s$) at 3.6 \micron. To investigate this, we compare our non-LTE calculations at the highest density regime (i.e., at LTE) with that of \S 4.1, i.e., calculations with all the transitions used in \citet{DB96}. We then define limitations of parameter ranges to guarantee a calculation error within $\sim$20 \%. These are: an upper limit of $T$ of 2000--3000 K for isothermal models; and a lower limit of $s$ of 3.0 and 4.7 for NY08 models with 49 and 36 energy levels, respectively (Appendix A). The errors for the other wavelengths (4.5, 5.8, and 8.0 \micron) are much smaller, allowing us to model the emission with sufficient parameter ranges for our interest.

\begin{figure*}
\epsscale{2}
\plotone{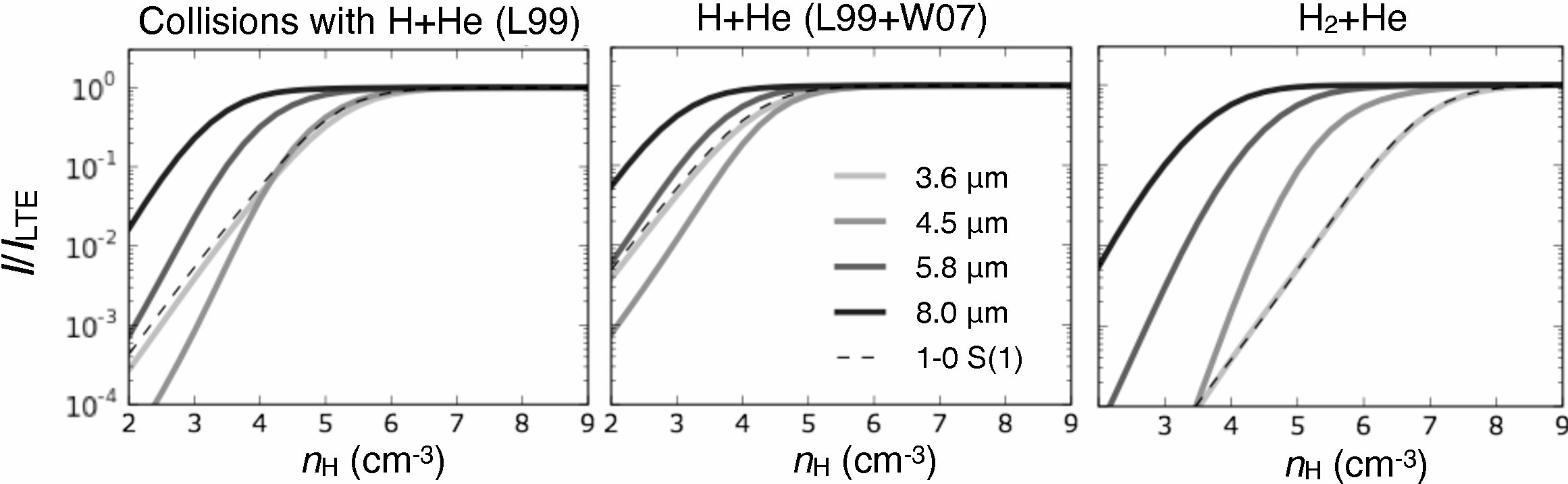}
\caption{IRAC flux of non-LTE H$_2$ at $T$=2000 K as a function of hydrogen number density. (left) For collisions with H+He, with the collisional rate coefficients of Le Boutlot et al. (1999) for both particles; (middle) Same as the left figure but the collisional rate coefficients of Wrathmall et al. (2007) are used for collisions with H; (right) For collisions with H$_2$ and He. All the fluxes are normalized to those at LTE. \label{fig_nLTE_curve}}
\end{figure*}

Figure \ref{fig_nLTE_curve} shows the IRAC and 1-0 S(1) fluxes as a function of hydrogen number density for three cases: (1) collisions with H+He, adopting the collisional rate coefficients of \citet{LeBourlot99}; (2) same as (1) but adopting the collisional rate coefficient for H of \citet{Wrathmall07}; and (3) collisions with H$_2$+He , adopting the collisional rate coefficients of \citet{LeBourlot99}. These calculations are made at temperature $T$=2000 K. For all cases the emission at shorter wavelengths requires higher densities for thermalization. The 5.8 \micron~ emission reaches LTE at higher density than 8.0 \micron; 3.6 and 4.5 \micron~ emission reaches LTE at even higher density. The 3.6 \micron~ emission reaches thermal equilibrium at a much higher density than 4.5 \micron~ emission for H$_2$+He collisions, while the difference between two lines are rather marginal for H+He collisions. The dependence of the 3.6 \micron~ emission on density is almost identical with 1-0 S(1). The non-LTE models with H+He collisions best explain the the observed morphological similarities between 3.6 \micron, 4.5 \micron, and 1-0 S(1) emission shown in \S 3.1.

The above discussions are often made with critical densities, although there are several difficulties for applying this to IRAC emission. These are:- (1)  each band includes multiple lines with different critical densities; and (2) ro-vibrationally excited states have two critical densities, representative of the thermalization of rotational and vibrational states, respectively. We thus alternatively define the critical density of IRAC  band emission as $I(n_{crit}) = 0.5 \times I_{LTE} $. This definition is based on the fact that the equation yields the correct critical density for the 2-level model.

Table \ref{tbl_n_crit} shows the critical densities of H$_2$ emission for four IRAC bands and 1-0 S(1) based on this definition ($T$=2000 K). As expected from Figure \ref{fig_nLTE_curve}, the densities are significantly higher at shorter wavelengths: the 5.8 \micron~ emission having a larger critical density than 8.0 \micron~ by a factor of 6--11; the 4.5 \micron~ having a higher critical density than the 5.8 \micron~ by a factor of 4--11. The critical densities for 3.6 and 4.5 \micron~ are similar to each other in case of H+He collisions, while the former is larger than the latter by a factor of 14--15 in case of H$_2$+He collisions. In addition to the H+He and H$_2$+He gas, we also show the critical densities without collisions with He. These show that collisions with He are not dominant compared with those with He or H$_2$, but not negligible, in particular for H$_2$+He gas (20--30 \%). Two different tables for H collisions \citep{LeBourlot99, Wrathmall07} yield critical densities different by a factor of 2--10. 

The critical densities for H+He and H$_2$+He collisions are significantly different, particularly at shorter wavelengths. The former is larger than the latter by by a factor of 50--500 for 3.6 \micron, and 3--6 for 8.0 \micron. This implies that the 3.6 \micron~ flux is determined by collisions with H even with a low H$_2$ dissociation rate  ($\gg$0.002--0.02). The 8.0 \micron~ flux is determined by collisions with H in cases the dissociation rate is much larger than 0.2--0.4. 

\begin{figure*}
\epsscale{2}
\plotone{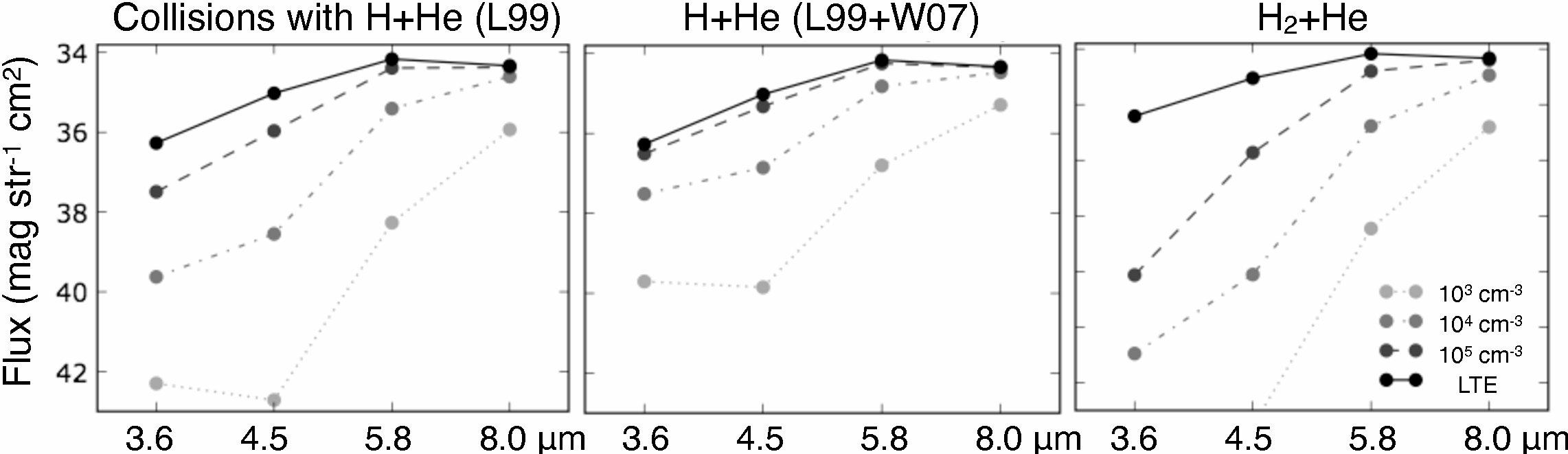}
\caption{Spectral energy distributions of non-LTE and LTE H$_2$ in IRAC bands at $T$=2000 K. (left) For collisions with H+He, with the collisional rate coefficients if Le Boutlot et al. (1999) for both particles; (middle) Same as the left figure but the collisional rate coefficients of Wrathmall et al. (2007) are used for collisions with H; (right) For collisions with H$_2$ and He. SEDs for non-LTE cases are shown for hydrogen number density of 10$^3$, 10$^4$, and 10$^5$ cm$^{-3}$. \label{fig_nLTE_SEDs}}
\end{figure*}

Figure \ref{fig_nLTE_SEDs} shows the SEDs of non-LTE and LTE gas with different particles and collisional rate coefficients.
Since the shorter wavelength IRAC bands require higher densities for thermalization, these are fainter in the SEDs at low densities.
As clearly seen in the figure, the deficit at lower densities is largest at 4.5 \micron~ for collisions with H+He, while it is largest at 3.6 \micron~ for collisions with H$_2$+He. In any case, the SEDs do not show 4.5 \micron~ excess observed in type-B SEDs.

\subsubsection{Color-color diagrams}
Figure \ref{fig_cc_nLTE1} shows color-color diagrams for the observed SEDs, and modeled emission for isothermal cases. 
The [4.5]--[5.8] vs [3.6]--[4.5] diagrams show that the colors of SEDs with weak/no 4.5-\micron~ excess are entirely explained either with (1) H+He collisions based on collisional rate coefficients by \citet{LeBourlot99}, or (2) H$_2$+He collisions. In the former case, the different colors observed in these SEDs ($\Delta m_{4.5} < 0.3$) are attributed to different densities but with a single temperature ($T \sim 2000$ K). In the latter case, for H$_2$+He collisions, the variation of the observed colors are attributed to a combination of different density and temperature in the case of H$_2$+He collisions. These models indicate a lower limit of hydrogen number density of $\sim$10$^5$ cm$^{-3}$ for H+He collisions, and $\sim$10$^7$ cm$^{-3}$ for H$_2$+He collisions. The H collisions based on parameters by \citet{Wrathmall07} explain only a fraction of the SEDs observed ([4.5]--[5.8]=0.0--1.2). None of the non-LTE models with an isothermal temperature structure explain the observed colors plotted in the [5.8]--[8.0] vs [4.5]--[5.8] diagrams.

Figure \ref{fig_cc_nLTE2} shows color-color diagrams for the observed SEDs, and modeled emission based on NY08. As in the isothermal cases,  
the [4.5]--[5.8] vs [3.6]--[4.5] diagrams show that the colors of SEDs with weak/no 4.5-\micron~ excess are explained with either H+He collisions based on \citet{LeBourlot99}, or H$_2$+He collisions. In both cases, the variation of the observed colors is attributed to a combination of different densities and power indexes $s$. In the [5.8]--[8.0] vs. [4.5]--[5.8]  diagrams, the observed colors are well fitted by the models at LTE. This is in contrast to the fact that the observations are explained by non-LTE regimes in the [4.5]--[5.8] vs [3.6]--[4.5] diagrams. This discrepancy is attributed to different critical densities between the four bands. As for the isothermal models,  the [4.5]--[5.8] vs [3.6]--[4.5] diagrams indicates a lower limit for the hydrogen number density of $\sim$10$^5$ cm$^{-3}$ for H+He collisions, and $\sim$10$^7$ cm$^{-3}$ for H$_2$+He collisions. Such densities yield close to LTE values in the [5.8]--[8.0] vs. [4.5]--[5.8]  diagrams.

\section{Discussion}
We summarize and discuss how the observed morphologies and SEDs are explained with models of H$_2$ emission, and implications for the physical conditions. Discussions of the morphologies and two types of SEDs are shown in \S 5.1--5.3, respectively. In \S 5.4, we discuss the applicability and limitations of thee-color images, which have been used by many authors.

\subsection{Observed Morphologies vs. H$_2$ Models}
As shown in \S 3.1, the morphologies observed at 3.6 \micron,  4.5 \micron~ and 1-0 S(1) match well, while the 5.8 and 8.0 \micron~ emission show slightly different morphologies. In regions where we can clearly see the morphological differences, the former is associated with the head of a bow shock or they exhibit clumpy/pachy structures; the latter is enhanced at the wakes of bow shocks or they exhibit more blurred structures. As discussed in \S 4.1.1 and \S 4.2.1, the similar morphologies of 3.6 \micron,  4.5 \micron~ and 1-0 S(1) emission is best explained by non-LTE models with H+He collisions. Since the collisional rate coefficients for H are much larger than H$_2$, it would require only a small dissociation rate.
As discussed in \S 4.2.1, a dissociation rate of $\gg$0.002--0.02 would allow collisional excitation and de-excitation dominated by H at $T$=2000 K, i.e., a typical temperature observed in Figure \ref{fig_cc_nLTE1}.

The morphological differences between 4.5, 5.8, and 8.0 \micron~ emission are attributed to either different temperatures or densities. In bow shocks, higher excitation is expected at the head than the wakes due to different shock velocities \citep[e.g.,][]{Hartigan87, Morse93a, Allen93}. The morphological differences in IRAC emission observed in such shocks are attributed to the fact that the thermal H$_2$ emission at shorter wavelengths tends to be associated with regions at higher temperatures (\S 4.1.1, \S 4.2.2). The fact that the emission at shorter wavelengths is enhanced in patchy structures may be due to different densities, rather than temperatures. This agrees with our calculations, which show that emissions at shorter wavelengths  have higher critical densities (\S 4.2.1). 

\subsection{Type-A SEDs vs. H$_2$ Models}
As shown in \S 3.2, the observed SEDs are categorized into: (A) those in which the flux monotonically increases with wavelength; and (B) those with excess emission at 4.5 \micron. As discussed in \S 4.2.2,  SEDs with no or weak 4.5-\micron~ excess (type-A or weak type-B) are explained by H$_2$ emission, in particular with simple shock models with a power-law cooling ($\Lambda \propto T^s$). Either collisions with H+He or H$_2$+He can explain the observed colors, but the former required the collisional rate coefficients of H of \citet{LeBourlot99}, rather than more recent calculations of \citet{Wrathmall07}. The color-color diagrams show that the IRAC emissions at short wavelengths (in particular 3.6 \micron) originate from H$_2$ in non-LTE, while that at long wavelengths (in particular 5.8 and 8.0 \micron) is due to H$_2$ at LTE. This discrepancy is attributed to different critical densities between four IRAC bands (\S 4.2.2).

\citet{Wrathmall07} state that they provide improved results compared with \citet{LeBourlot99}. However, models with their coefficients explain only a fraction of observed IRAC colors at short wavelengths (\S 4.2.2 and Figure \ref{fig_cc_nLTE2}). If their coefficients are accurate, the observed SEDs may be explained as follows. Thermal collisions are dominated by H$_2$+He for SEDs with steep gradients ([4.5]--[5.8]$>$1.2), but are dominated by H(+He) for SEDs with shallow gradients ([4.5]--[5.8]$<$1.2). This explanation is consistent with the fact that a higher dissociation rate is expected in regions at higher temperatures.

If we assume H$_2$+He collisions, Figure \ref{fig_cc_nLTE2} implies a gas density significantly higher than that provided by \citet{Neufeld09}, who estimated the density in several outflows with low-$J$ lines assuming H$_2$+He collisions. In Figure \ref{fig_cc_nLTE2}, observed SEDs with weak or no 4.5-\micron~ excess ($\Delta m_{4.5} < 0.3$) are explained with a density of more than 10$^7$ cm$^{-3}$. In contrast, Neufeld et al. (2009) fit their spectra with much smaller densities ($\sim$10$^4$ cm$^{-3}$). This discrepancy is consistent with the fact that our selection of positions is biased to relatively bright regions in all four IRAC bands. Due to a high critical density at short wavelengths (10$^6$--10$^7$ cm$^{-3}$ for 3.6 and 4.5 \micron~ emission for H$_2$+He collisions), such positions should be associated with gas at high densities. This selection bias would also explain why a different power-law index $s$ is required to explain our SEDs (3.0--6.0) compared to the low-$J$ spectra observed by \citet{Neufeld09} (2--3).

The temperature structure we used ($\Lambda \propto T^s$ at $T_{max}$=4000 K) corresponds to a simplified $J$-shock, although our calculations would not reject $C$-shocks for explaining the type-A SEDs. Indeed, the isothermal models also account for the observed colors in the [4.5]--[5.8] vs. [3.6]--[4.5] diagrams, as shown in Figures \ref{fig_cc_LTE} and \ref{fig_cc_nLTE2}. This suggests that the temperature structure near the highest temperature, which affects the colors at short wavelengths, is arbitrary to some extent. Thus, a flat-top temperature structure predicted by $C$-shock models \citep[e.g.,][]{DRD83, Kaufman96a, LeBourlot02} may also fit well the type-A SEDs. Testing this scenario is beyond the scope of this paper.

\subsection{Type-B SEDs}
Our results suggest that SEDs with a large 4.5-\micron~ excess (strong type-B) cannot be solely attributed to thermal H$_2$ emission. Modeled SEDs at either LTE and non-LTE do not show such excess emission (\S 4.1.1 and \S 4.2.1). The [5.8]--[8.0] vs. [4.5]--[5.8] diagrams in Figures \ref{fig_cc_LTE} and \ref{fig_cc_nLTE2} show discrepancies between the observed SEDs with $\Delta m_{4.5} > 0.3$ and models up to [4.5]--[5.8]$\sim$1. These discrepancies between the strong type-B SEDs and H$_2$ models are in contrast to the analysis by YL09, who explain such colors in shocks associated with HH 54 ([3.6]--[4.5] $>$1.0, [4.5]--[5.8] $<$0.7) with non-LTE H$_2$ calculations with the lowest 49 levels. However, the temperature YL09 measured is 2000--4000 K, implying that the [3.6]--[4.5]  color they model would significantly suffer from errors due to the limited number of transitions (up to $\sim$0.6 mag, see Appendix A).

Thus, we suggest a different origin, or the presence of an additional emission component for the type-B SEDs. The continuous color variation shown in Figure \ref{fig_obs_excess_corr} suggests continuous variation in density and temperature in the observed regions.
The contamination of CO vibrational emission, suggested by several authors \citep{Noriega-Crespo04, Reach06, Qiu08}, is the most likely candidate to explain all the observed tendencies. The entire type-B SEDs, which show (1) excess emission at 4.5-\micron; and (2) shallower magnitude gradient at the other wavelengths (Figures \ref{fig_obs_SEDs} and  \ref{fig_obs_excess_corr}), are explained with a combination ofΠdensity and temperature higher than type-A SEDs. The former is explained with the fact that the CO emission contributes to mostly the 4.5-\micron~ flux \citep[see][]{Reach06}, and that the contribution is comparable to H$_2$ emission if the number density of H atoms is 10$^7$--10$^8$ cm$^{-3}$ (see Figure 11 of NY08). The latter is explained by the fact that the SEDs for thermal H$_2$ emission show shallower gradients in magnitude at a high density or temperature.

Alternative candidates for the 4.5-\micron~ excess emission could be H I (in particular Br $\alpha$) and [Fe II] \citep[][YL09]{Noriega-Crespo04, Reach06, Qiu08}. Both emission mechanisms are expected in dissociative shocks \citep[e.g.,][]{HM89}. \citet{HM89} show that such shocks diminish thermal H$_2$ emission at $T>1000$ K, since reformation of most of the H$_2$ molecules occurs at lower temperatures. This trend is clearly seen in bow shocks in the Orion-KL outflow by \citet{Tedds99}, who show that the [Fe II] 1.64 \micron~ emission is associated with the head, while the H$_2$ 2.12 \micron~ emission is associated with the wakes. Such a deficit of the observing flux is not observed in the type-B SEDs: these are comparable to or brighter in all of the four IRAC bands than type-A SEDs. We thus conclude that the excess emission at 4.5-\micron~ is not due to either H I or [Fe II].

Based on IRAC observations, \citet{SmithH06} reported the presence of UV fluorescent H$_2$ emission in shocks associated with the DR 21 outflow. To investigate this possibility, we calculated the flux of photodissocoiation regions (PDRs) modeled by \citet{DB96} with a variety of densities ($n_H=10^2$ to 10$^6$ cm$^{-3}$) and UV field ($\chi = 1$ to 10$^5$). These are shown in the color-color diagrams in Figure \ref{fig_cc_vsPDRs}, together with the observed colors in shocks. The observed colors show no correlation with the PDR models, implying that contribution of UV fluorescent H$_2$ emission to our SEDs is negligible.

\subsection{Discriminating Shocks from PAH and Stars with IRAC Colors}
As described in \S 1,  three-color images (blue, green, red for 3.6, 4.5 and 8.0 \micron, respectively)  have been used by a number of authors to discriminate between stars, PAH, shocks and scattered continuum in star forming regions. Stars are relatively bright at shorter wavelengths hence these appear blue in such images; PAH emission has a remarkable excess at 8.0 \micron~ hence it appears red; and the others tend to appear green.

Figure \ref{fig_cc_RGB}  shows the 3.6, 4.5, and 8.0 \micron~ colors observed in shocks, PAH in NGC 7023 estimated using ISO spectra (IRAC Data Handbook), and stars. The figure shows that some shocks can look "red" in the three-color images. As shown in \S 4, such shocks have relatively low temperature and/or density. The figure also shows that the [3.6]--[4.5] color allows us to clearly discriminate shocks ([3.6]--[4.5]$>$1.1) from stars and PAH ([3.6]--[4.5]$<$1.1). The observed [3.6]--[4.5] color is lowest for the SEDs with [4.5]--[8.0]=1.0--2.0, and larger [3.6]--[4.5] colors for the other SEDs are due to either their low densities (\S 4.2.2) or contaminating emission at 4.5 \micron~ (\S 5.1). 
The presence of a lower limit in the [3.6]--[4.5] color is also supported by empirical use of 4.5/3.6-\micron~ flux ratio to identify shocks in some recent studies \citep{Qiu08, Zhang09}. 

Figure \ref{fig_L1448S} shows the three-color image and 4.5/3.6-\micron~ flux ratio in the southern end of  the L 1448 jet. The brightness scale of the three-color image is automatically adjusted using ds9 developed by Smithsonian Astrophysical Observatory. The figure shows that this shock region contain both greenish and reddish components, agreeing with the above discussion with Figure \ref{fig_cc_RGB}. The flux ratio map shows that the ratio is comparable or higher than $\sim 1.5$ (corresponding to [3.6]--[4.5] $>$ 0.9) in the entire jet, in contrast to the stars in the same image (flux ratio $\ll$ 1.5).
 
YL09 state the possibility of PAH emission in shocks seen in IRAC images. Our results do not show clear evidence for such contaminating emission. In Figure \ref{fig_cc_RGB}, the colors measured in shocks are distributed perpendicularly to the color of PAH, with a significant color offset.  

\section{Summary and Conclusions}

We present detailed analysis of shocked emission in four Spitzer-IRAC bands associated with six Herbig-Haro objects (HH 212, L1448, L1157, HH 211, BHR 71 and HH 54). Our research goal is to fully investigate the nature of this shocked emission: i.e., is this explained by emission from molecular hydrogen; what are the implications for temperature and density of these regions; what are the possible sources of contaminating emission. We compare the observed morphologies and flux ratios measured at high-S/N regions with models of thermally excited molecular hydrogen emission with different excitation conditions and two types of temperature structures: isothermal and power-law cooling ($\Lambda \propto T^s$). For HH 212, L 1448 and HH 211, narrow-band images of 1-0 S(1) observed using Subaru-MOIRCS and VLT-ISAAC are added for our study.

The morphologies observed at 3.6 \micron~ and  4.5 \micron~ and 1-0 S(1) match well, while the 5.8 and 8.0 \micron~ emission show different morphologies. In regions where we can clearly see the morphological differences, the former is either associated with the head of bow shock or they exhibit clumpy/pachy structures; the latter is either enhanced at the wakes of bow shocks or they exhibit more blurred structures. The observed SEDs in selected regions are categorized into: (A) those in which the magnitude monotonically increases with wavelength; and (B) similar to (A) but with excess emission at 4.5 \micron. The type-B SEDs tend to be brighter than type-A in each object, and the magnitude of excess is well correlated with the color of the entire SED.

We compare the above results with (1) LTE calculations with 245 energy levels, including 207 lines in four IRAC bands; and (2) non-LTE calculations with 36--49 levels, including 32--45 lines in four IRAC bands. The non-LTE calculations are made for collisional particles for two cases (H+He and H$_2$+He). These show that H$_2$ emission explains the morphological similarities between 3.6 \micron~ and  4.5 \micron~ and 1-0 S(1), and their differences from 5.8  and  8.0 \micron~ emission. The former fact is particularly well explained with non-LTE H$_2$ if the dissociation rate is much larger than 0.002--0.02, allowing thermal collisions dominated by atomic hydrogen. The latter fact is explained by the fact that H$_2$ emission at longer wavelength bands should be enhanced at lower temperatures and/or densities. 

The H$_2$ models also successfully explain type-A SEDs. The models with power-law cooling rate explain different colors between SEDs, with different densities and power indexes particularly well. These are explained by collisions with either H$_2$+He, or H+He with collisional rate coefficients of H provided by \citet{LeBourlot99}. The variation in SEDs shows that the colors at short wavelengths are attributed to non-LTE H$_2$, while those at long wavelengths are attributed to LTE H$_2$ due to their lower critical densities. In contrast to the type-A SEDs, our calculations suggest that thermal H$_2$ cannot explain the 4.5-\micron~ excess in type-B SEDs. We suggest that the combination of CO and H$_2$ emission qualitatively explains the observed tendencies. 
It is not likely that H I, [Fe II], fluorescent H$_2$ or PAH significantly contributes to the SEDs we observed.

Three-color images (blue, green, red for 3.6, 4.5 and 8.0 \micron, respectively) are widely used by many authors to instantly discriminate between shocks, PAH emission and stars. We show that shocked emission at low temperatures (i.e., large power-law index for cooling function) can be ``red'' in such images, analogous to PAH emission. Our new and thorough measurements of colors suggest that these can be discriminated with the [3.6]--[4.5] color: $\gtrsim$1.0 for shocks and $\lesssim$1.0 for PAH.



\acknowledgments

We thank the anonymous referee for useful comments.
We are grateful for Dr. Ohyama for useful discussions about data analysis. The IRAC images were obtained through the Spitzer archive operated by IPAC.
The narrow-band images of the H$_2$ 2.12 \micron~ emission for HH 212 were obtained through the ESO archive operated by European Southern Observatory.
MT acknowledges support from National Science Council of Taiwan (Grant No. 97WIA0100327).
The research made use of the Simbad data base operated at CDS, Strasbourg, France, and the NASA's Astrophysics Data System Abstract Service.



{\it Facilities:} \facility{Spitzer Space Telescope (IRAC)}, \facility{Subaru (MOIRCS)}, \facility{VLT (ISAAC)}.



\appendix

\section{Error of non-LTE calculations with limited number of transitions}

Our non-LTE calculations of thermal H$_2$ include the lowest 49 levels ($E_u/k < 2 \times 10^4$ K)  for thermal collisions with H or H+He , and the lowest 36 levels  ($E_u/k < 1.69 \times 10^4$ K) for collisions with H$_2$ or H$_2$+He. 
In practice, the transitions from higher energy levels contribute to the total IRAC flux, thus our non-LTE calculations would have larger errors in particular at short wavelengths and high temperatures. 

\begin{figure*}
\epsscale{2}
\plotone{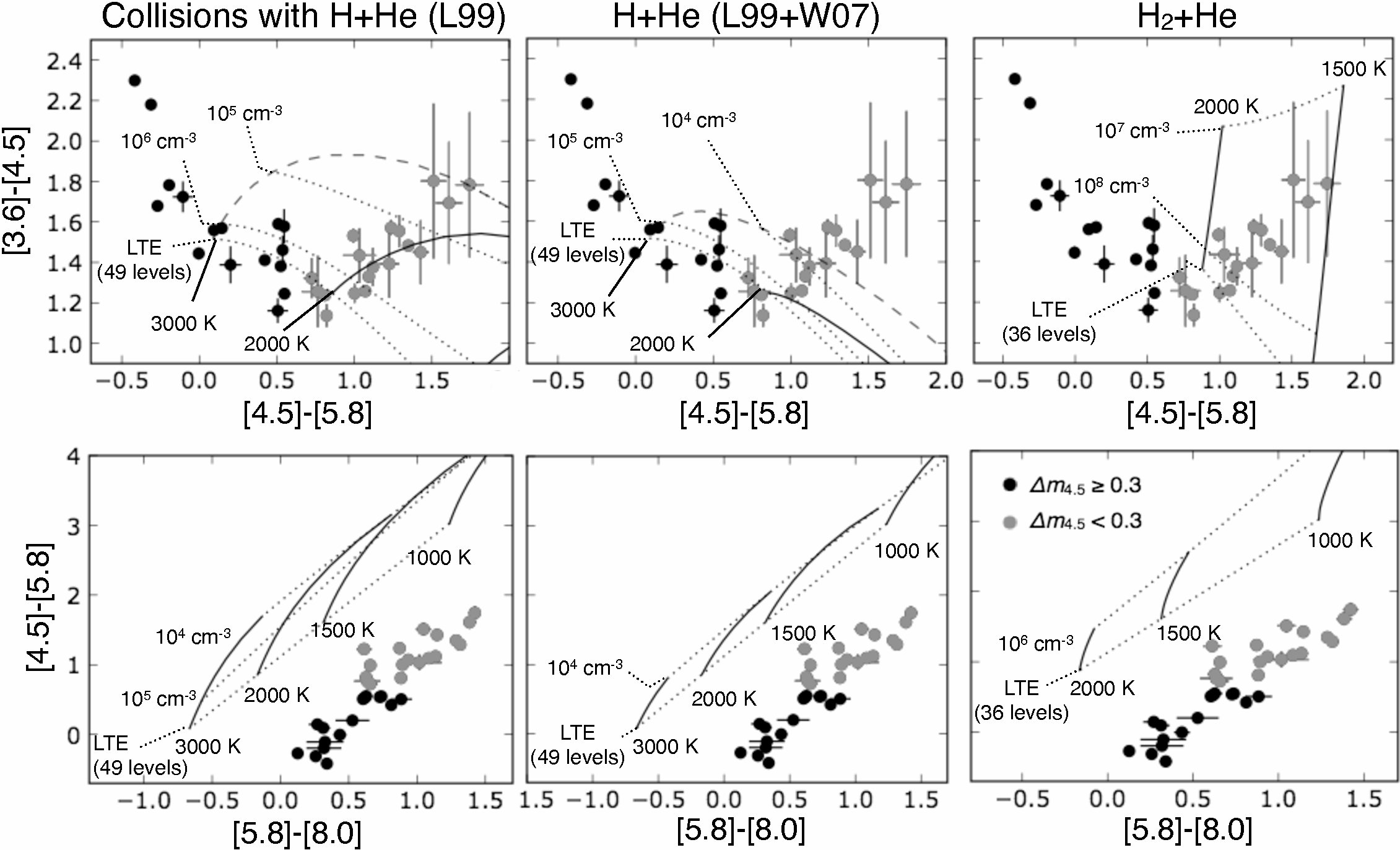}
\caption{Color-color diagrams for observations and non-LTE models for isothermal cases.
(left) For collisions with H+He, with the collisional rate coefficients by Le Boutlot et al. (1999) for both particles; (middle) Same as the left figure but the collisional rate coefficients by Wrathmall et al. (2007) are used for collisions with H; (right) For collisions with H$_2$ and He. The upper and lower plots are diagram with [4.5]--[5.8] vs. [3.6]--[4.5] and [5.8]--[8.0] vs. [4.5]--[5.8], respectively. In the upper-left and upper-middle figures, the modeled colors for $T$=3000 K are shown in dashed line, since it would have a relatively large error for calculations ($>$0.2 mag. for [3.6]-[4.5], see Appendix A).
 \label{fig_cc_nLTE1}}
\end{figure*}

\begin{figure*}
\epsscale{2}
\plotone{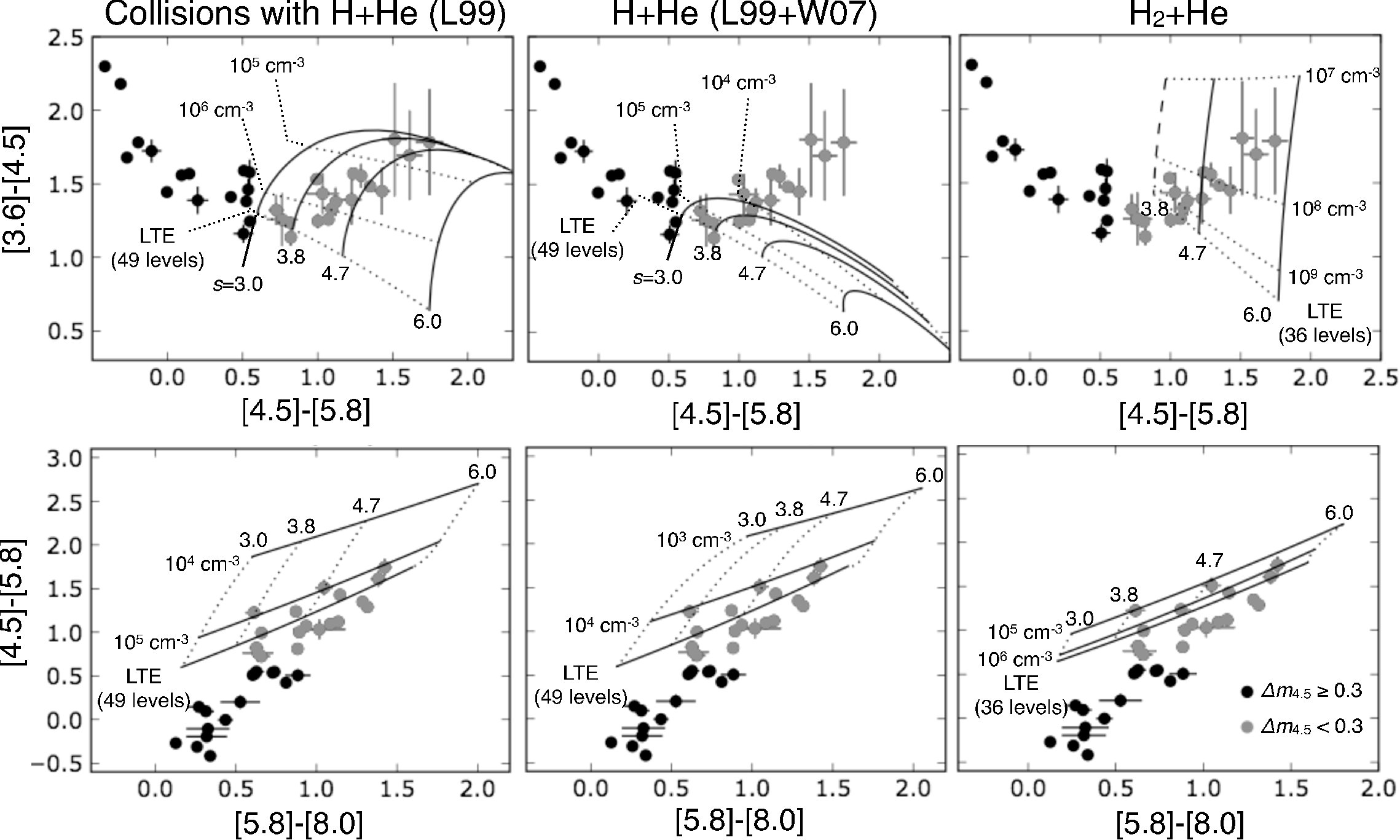}
\caption{Same as Figure \label{fig_cc_LTE1} but with temperature structures used in Neufeld \& Yuan (2008). Solid and dotted lines in the upper figure shows the modeled values with constant power indexes $s$ and densities, respectively. In the lower figures, the solid lines are used for models with constant densities to clearly show that the LTE values match the observations. In the upper-right figure modeled colors at $s$=3.8 is shown in dashed lines due to a relatively large uncertainty ($>$0.2 mag. for [3.6]-[4.5], see Appendix A).
\label{fig_cc_nLTE2}}
\end{figure*}

\begin{figure*}
\epsscale{1.4}
\plotone{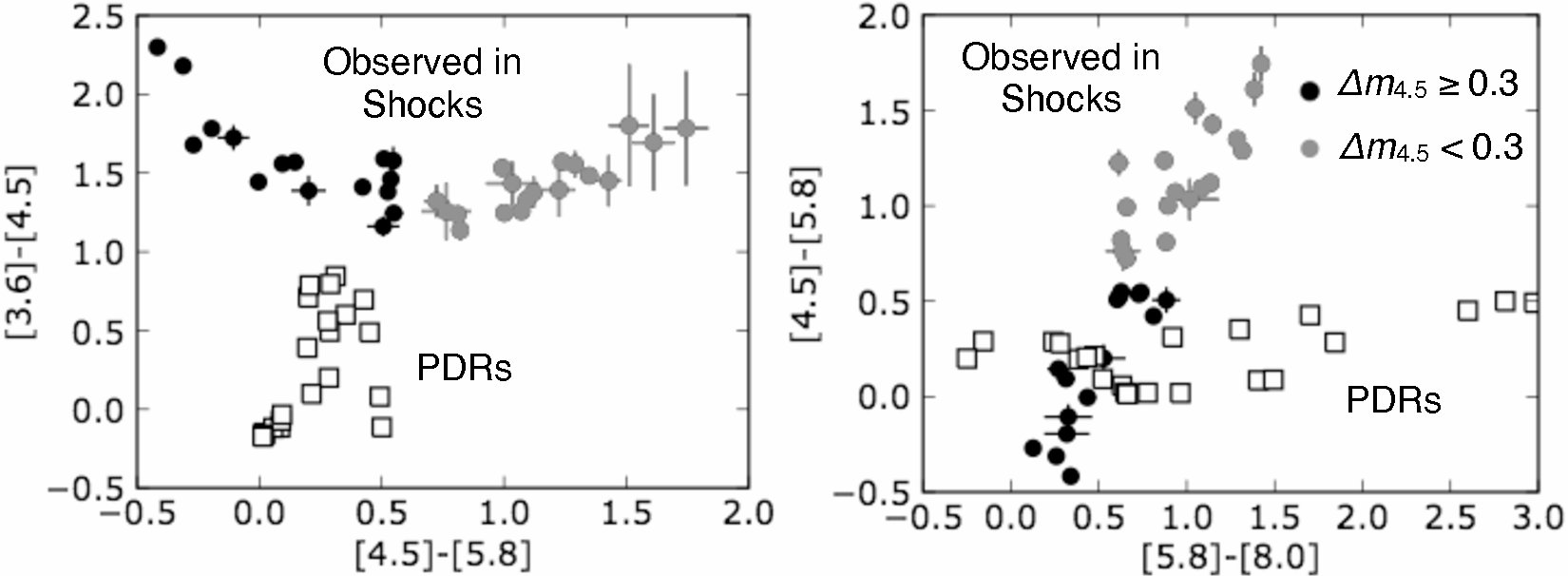}
\caption{Color-color diagrams for SEDs observed in shocks (black and gray dots), and PDR models provided by Draine \& Bertoldi (1996) at $n_H = 10^2$ to 10$^6$ cm$^{-3}$ and $\chi= 1$ to $10^5$ (squares). \label{fig_cc_vsPDRs}}
\end{figure*}

\begin{figure*}
\epsscale{1.4}
\plotone{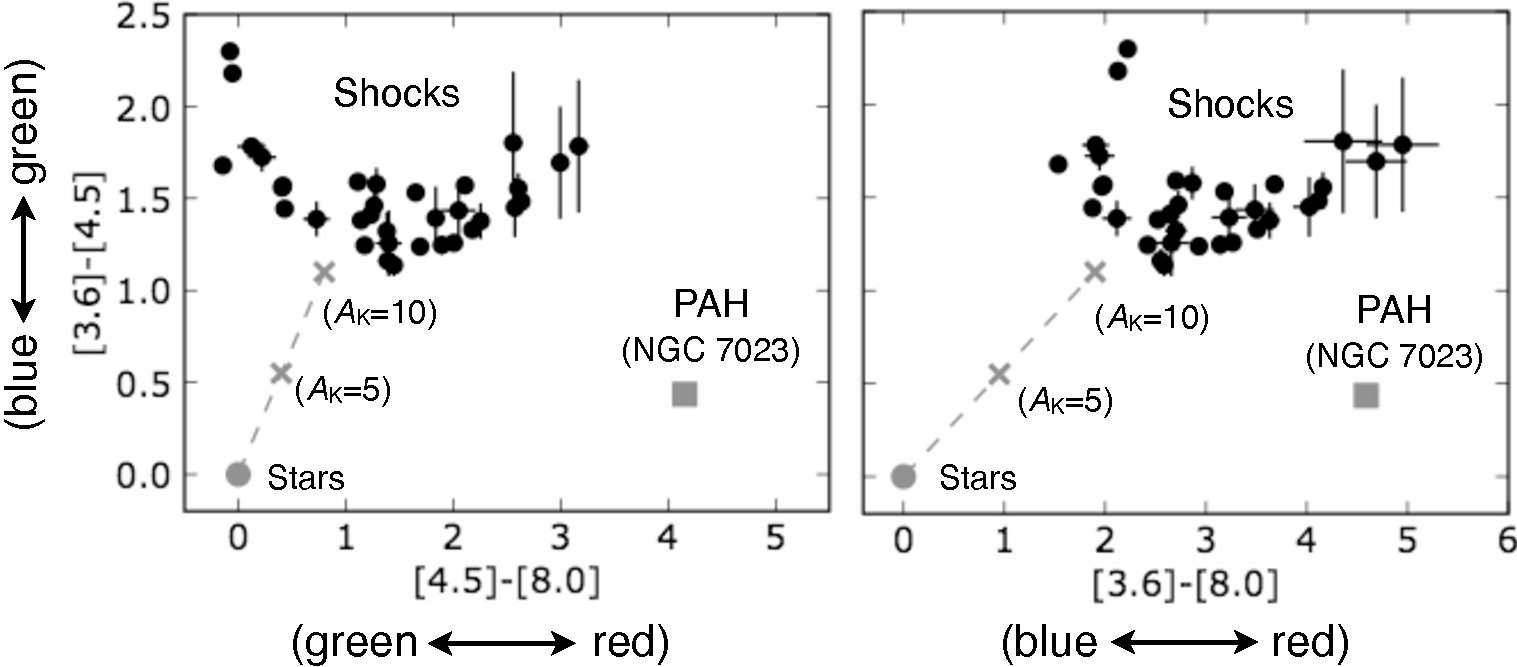}
\caption{3.6-, 4.5-, and 8.0-\micron~ colors of shocks, stars, and PAH. The corresponding color for the three-color images are also shown beside the labels for the horizontal and vertical axes. Extinction added to the stars is based on Chapman et al. (2009) for $A_K>2$. Color of PAH in NGC 7023 is based on IRAC Data Handbook. \label{fig_cc_RGB}}
\end{figure*}

\begin{figure}
\epsscale{1.0}
\plotone{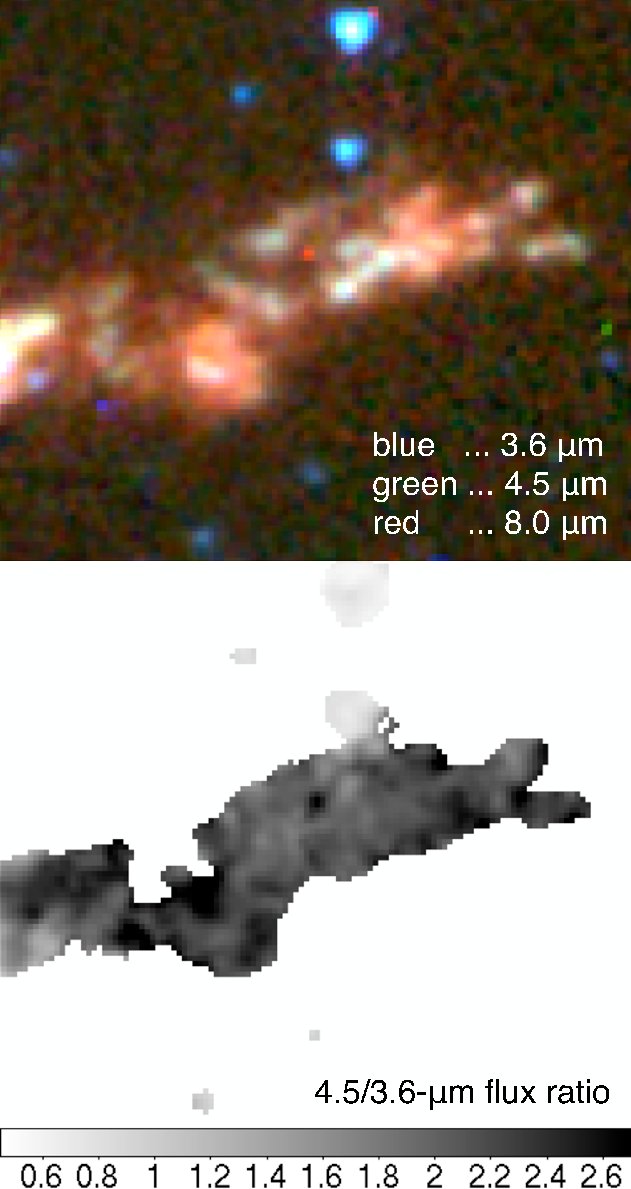}
\caption{(left) Three-color image at the south end of L 1448. The color scale is automatically adjusted using ds9, developed by Smithsonian Astrophysical Observatory. (right) The 4.5/3.6-\micron~ flux ratio. The measurements are made at positions where the signal-to noise exceeds 10-$\sigma$ both at 3.6 and 4.5 \micron. Before taking the flux ratio, the 3.6-\micron~ image is  convolved with the PRF at 4.5 \micron, and 4.5-\micron~ image is  convolved with the PRF at 3.6 \micron, to develop higher signal-to-noise. \label{fig_L1448S}}
\end{figure}

\begin{figure*}
\epsscale{1.4}
\plotone{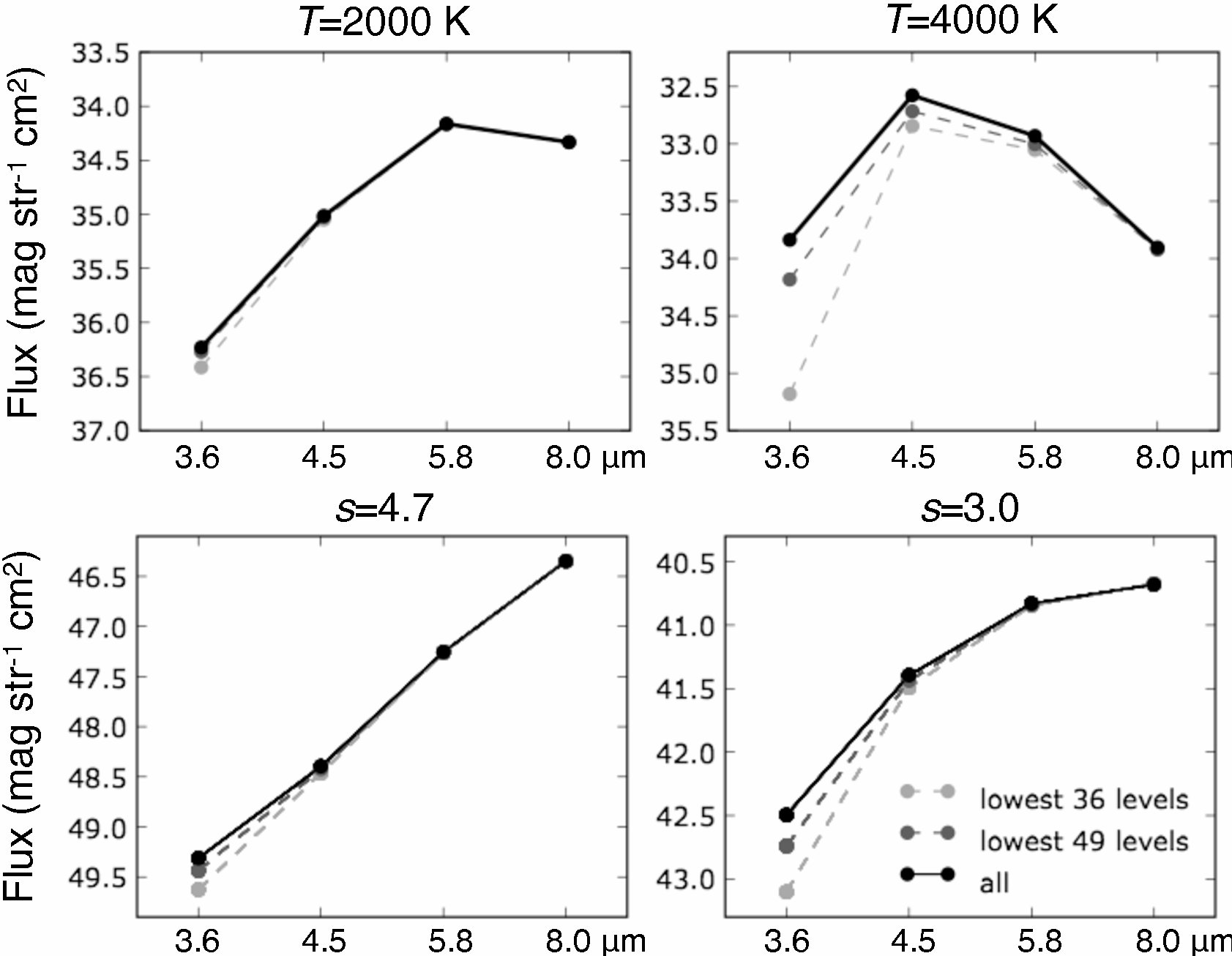}
\caption{IRAC flux of LTE H$_2$ calculated with the lowest 36/49 levels, and all the transitions included in Draine \& Bertoldi (1996). (upper left) For the isothermal case at $T$=2000 K; (upper right) same but $T$=4000 K; (lower left) Neufeld \& Yuan (2008) model with $s$=4.7; (lower right) same but $s$=3.0. \label{fig_nLTE_Tsmax}}
\end{figure*}

To investigate the missing flux from the upper levels,  we calculate the fraction of the LTE flux from the lowest 36--49 levels, and compare them with the LTE flux we derive in \S 4.1.1. Figure \ref{fig_nLTE_Tsmax} shows the difference of SEDs in these cases, at different temperatures and power indexes for cooling.
Tables \ref{tbl_nLTE_Tmax1} shows the contribution of the lowest 36--49 levels to the total flux at different temperatures, and different power indexes for the cooling function ($s$). These show that the 3.6-$\micron$ flux would significantly suffer from errors due to  limited numbers of 
transitions. The lowest 36--49 levels are esponsible for only 20--50 \% of the total flux at $T$=4000 K; 57--80 \% for $s$=3.0. Table \ref{tbl_nLTE_Tmax2} shows the estimated errors for IRAC colors in magnitudes. The parameter range of non-LTE calculations in Figures \ref{fig_cc_nLTE1} and \ref{fig_cc_nLTE2} are determined based on these tables.

Note that the population of the lower 36--49 levels is still responsible for a large fraction of population of H$_2$ (94--97 \% at $T$=4000 K). The significant contribution from the upper-level transitions to the 3.6-$\micron$ flux is due to a large number of lines ($\sim$100) and their large $A$ coefficients.

%
%






\clearpage





\clearpage


\clearpage

\clearpage

\clearpage

\clearpage

\clearpage


\clearpage

\clearpage

\clearpage


\clearpage







\clearpage


\begin{table}
\begin{center}
\caption{Observed IRAC Flux \label{tbl_obs_IRAC}}
\begin{tabular}{ccccccccccccc} \tableline \tableline
Object & Position &\multicolumn{11}{c}{Flux (mag arcsec$^{-2}$)} \\ 
&& \multicolumn{2}{c}{[3.6]} && \multicolumn{2}{c}{[4.5]} && \multicolumn{2}{c}{[5.8]} && \multicolumn{2}{c}{[8.0]} \\ \hline
 HH 212  &   1 & 18.60 & ($<$0.03) && 16.82 & ($<$0.03) && 17.01 & ($<$0.03) &&  16.69 & (~0.12~) \\
        &   2 & 18.75 & ($<$0.03) && 17.19 & ($<$0.03) && 17.09 & ($<$0.03) &&  16.78 & (~0.04~) \\
        &   3 & 17.80 & ($<$0.03) && 16.23 & ($<$0.03) && 16.09 & ($<$0.03) &&  15.81 & (~0.05~) \\
        &   4 & 20.27 & (~0.12~) && 18.83 & (~0.06~) && 17.80 & (~0.08~) && 16.78 & (~0.13~) \vspace{0.1cm} \\
 L 1448  &   1 & 17.96 & ($<$0.03) && 16.71 & ($<$0.03) && 16.16 & ($<$0.03) &&  15.54 & ($<$0.03) \\
        &   2 & 17.53 & ($<$0.03) && 16.15 & ($<$0.03) && 15.62 & ($<$0.03) &&  15.01 & ($<$0.03) \\
        &   3 & 19.51 & (~0.15~) && 18.06 & (~0.04~) && 16.63 & ($<$0.03) &&  15.49 & ($<$0.03) \\
        &   4 & 19.16 & (~0.09~) && 17.79 & ($<$0.03) && 16.67 & ($<$0.03) && 15.53 & ($<$0.03) \vspace{0.1cm} \\
 L 1157  &   1 & 17.52 & ($<$0.03) && 16.10 & ($<$0.03) && 15.68 & ($<$0.03) &&  14.87 & ($<$0.03) \\
        &   2 & 18.27 & ($<$0.03) && 17.04 & ($<$0.03) && 16.23 & ($<$0.03) &&  15.35 & ($<$0.03) \\
        &   3 & 19.35 & (~0.04~) && 18.10 & ($<$0.03) && 17.10 & ($<$0.03) &&  16.20 & ($<$0.03) \\
        &   4 & 19.81 & ($<$0.03) && 18.32 & ($<$0.03) && 16.98 & ($<$0.03) &&  15.69 & ($<$0.03) \\
        &   5 & 20.15 & (~0.07~) && 18.59 & ($<$0.03) && 17.31 & ($<$0.03) && 15.99 & ($<$0.03) \vspace{0.1cm} \\
 HH 211  &   1 & 16.80 & ($<$0.03) && 14.62 & ($<$0.03) && 14.93 & ($<$0.03) &&  14.68 & ($<$0.03) \\
        &   2 & 18.08 & (~0.03~) && 16.92 & (~0.04~) && 16.41 & (~0.04~) &&  15.53 & (~0.06~) \\
        &   3 & 17.53 & (~0.07~) && 15.80 & ($<$0.03) && 15.91 & (~0.05~) &&  15.58 & (~0.12~) \\
        &   4 & 17.77 & (~0.08~) && 16.38 & (~0.04~) && 16.18 & (~0.05~) && 15.65 & (~0.10~) \vspace{0.1cm} \\
 BHR 71  &   1 & 18.88 & (~0.15~) && 17.63 & (~0.08~) && 16.86 & (~0.05~) &&  16.23 & (~0.07~) \\
        &   2 & 17.69 & (~0.05~) && 16.56 & ($<$0.03) && 15.74 & ($<$0.03) &&  15.11 & ($<$0.03) \\
        &   3 & 16.63 & ($<$0.03) && 14.95 & ($<$0.03) && 15.22 & ($<$0.03) &&  15.09 & ($<$0.03) \\
        &   4 & 17.68 & ($<$0.03) && 15.38 & ($<$0.03) && 15.80 & ($<$0.03) &&  15.46 & ($<$0.03) \\
        &   5 & 17.43 & (~0.06~) && 15.97 & ($<$0.03) && 15.44 & ($<$0.03) &&  14.71 & ($<$0.03) \\
        &   6 & 17.68 & (~0.08~) && 16.10 & ($<$0.03) && 15.55 & ($<$0.03) &&  14.81 & ($<$0.03) \\
        &   7 & 17.44 & ($<$0.03) && 15.99 & ($<$0.03) && 16.00 & ($<$0.03) && 15.56 & (~0.04~) \vspace{0.1cm} \\
  HH 54  &   1 & 17.72 & ($<$0.03) && 16.39 & ($<$0.03) && 15.30 & ($<$0.03) &&  14.21 & ($<$0.03) \\
        &   2 & 17.36 & ($<$0.03) && 16.10 & ($<$0.03) && 15.03 & ($<$0.03) &&  14.10 & ($<$0.03) \\
        &   3 & 18.01 & (~0.03~) && 16.44 & ($<$0.03) && 15.21 & ($<$0.03) &&  14.33 & ($<$0.03) \\
        &   4 & 17.84 & ($<$0.03) && 16.31 & ($<$0.03) && 15.32 & ($<$0.03) &&  14.66 & ($<$0.03) \\
        &   5 & 17.71 & ($<$0.03) && 16.12 & ($<$0.03) && 15.62 & ($<$0.03) &&  15.01 & ($<$0.03) \\
        &   6 & 20.52 & (~0.35~) && 18.73 & (~0.08~) && 16.99 & (~0.04~) &&  15.57 & ($<$0.03) \\
        &   7 & 20.34 & (~0.29~) && 18.64 & (~0.07~) && 17.03 & (~0.04~) &&  15.65 & ($<$0.03) \\
        &   8 & 20.96 & (~0.37~) && 19.16 & (~0.06~) && 17.64 & (~0.04~) &&  16.60 & ($<$0.03) \\
        &   9 & 18.81 & (~0.09~) && 17.49 & (~0.03~) && 16.76 & (~0.03~) &&  16.11 & (~0.04~) \\
        &  10 & 19.40 & (~0.16~) && 18.01 & (~0.05~) && 16.79 & (~0.03~) && 16.17 & (~0.04~) \vspace{0.1cm} \\ \tableline
\end{tabular}
\end{center}
\end{table}

\clearpage


\begin{table}
\begin{center}
\caption{Observed 1-0 S(1) Flux \label{tbl_obs_10S1}}
\begin{tabular}{ccc} \tableline \tableline
Object & Position & Flux\\ 
             &                & (W m$^{-2}$ str$^{-1}$)\\  \tableline
 HH212  &   1  &  1.1$\times 10^{-6}$ \\
                &   2  &  7.1$\times 10^{-7}$ \\
                &   3  &  2.1$\times 10^{-6}$ \\
                &   4  &  2.7$\times 10^{-7}$ \vspace{0.1cm} \\
 L1448   &   1  &  2.9$\times 10^{-6}$\\
                &   2  &  4.1$\times 10^{-6}$\\
                &   3  &  1.0$\times 10^{-6}$\vspace{0.1cm} \\
 HH211  &   1  &  7.3$\times 10^{-6}$\\
                &   2  &  1.6$\times 10^{-6}$\\
                &   3  &  3.3$\times 10^{-6}$\\
                &   4  &  2.7$\times 10^{-6}$\vspace{0.0cm} \\  \tableline
\end{tabular}
\end{center}
\end{table}

\clearpage


\begin{table*}
\begin{center}
\caption{Summary of Transitions included in Le Bourlot et al.  (1999) and Wrathmall et al. (2007)  \label{tbl_L99W07_summary}}
\begin{scriptsize}
\begin{tabular}{llccl} \tableline \tableline
Paper & Particle & Counterpart  & Total energy & Completeness of Transitions\\ 
            &                & for collisions & levels included & \\ \tableline 
Le Bourlot (1999) & ortho-H$_2$ & H, He, para-H$_2$   & 24 & up to 23 levels ($E_u/k \le 19086$ K)\\
                                  &                         & ortho-H$_2$              & 22 & up to 22 levels ($E_u/k \le 18979$ K)\\
                                 & para-H$_2$ & H, He , para-H$_2$  & 27 & up to 26 levels ($E_u/k \le 19112$ K)\\
                                  &                          & ortho-H$_2$              & 19  & up to 19 levels ($E_u/k \le 16880$ K)\\
Wrathmall et al. (2007) & ortho-H$_2$ & H & 54 & up to 54 levels ($E_u/k \le 34037$ K)\\
                                          & para-H$_2$  & H & 54 & up to 54 levels ($E_u/k \le 32127$ K)\\ \tableline

\end{tabular}
\end{scriptsize}
\end{center}
\end{table*}


\begin{table*}
\begin{center}
\caption{Critical hydrogen nucleus densityin cm$^{-3}$ defined by $I/I_{LTE}$=0.5 ($T$=2000 K) \label{tbl_n_crit}}
\begin{tabular}{llccccc} \tableline \tableline
Collisions with & Reference\tablenotemark{a}  & \multicolumn{4}{c}{IRAC Band} & 1-0 S(1)\\
& & [3.6] & [4.5] & [5.8] & [8.0] \\ \tableline
H                  & W07  &$2.3\times 10^{4}$ & $3.6 \times 10^{4}$ & $8.7 \times 10^{3}$ & $1.5 \times 10^{3}$ &$1.7 \times 10^{4}$\\
H+He          & W07 (H) + L99 (He) &$2.3\times 10^{4}$ & $3.6 \times 10^{4}$ & $8.5 \times 10^{3}$ & $1.4 \times 10^{3}$ &$1.7 \times 10^{4}$\\
H                  & L99  &$2.2\times 10^{5}$ & $1.5 \times 10^{5}$ & $2.3 \times 10^{4}$ & $3.4 \times 10^{3}$ &$1.6 \times 10^{5}$\\
H+He          & L99  &$2.2\times 10^{5}$ & $1.4 \times 10^{5}$ & $2.1 \times 10^{4}$ & $3.0 \times 10^{3}$ &$1.6 \times 10^{5}$\\
H$_2$        & L99   &$1.8\times 10^{7}$ & $1.2 \times 10^{6}$ & $1.1 \times 10^{5}$ & $1.0 \times 10^{4}$ &$1.7\times 10^{7}$\\
H$_2$+He & L99   &$1.2\times 10^{7}$ & $8.8 \times 10^{5}$ & $8.0 \times 10^{4}$ & $7.8 \times 10^{3}$ &$1.1\times 10^{7}$\\ \tableline
\end{tabular}\\
\tablenotetext{a}{W07 ... Wrathmall et al. (2007); L99 ... Le Bourlot et al. (1999)}
\end{center}
\end{table*}


\begin{table*}
\begin{center}
\caption{Contribution of the lowest 36 and 49 levels to the total LTE flux (per cent)\label{tbl_nLTE_Tmax1}}
\begin{tabular}{ccccccccccc} \tableline \tableline
      && \multicolumn{4}{c}{Lowest 36 Levels}  && \multicolumn{4}{c}{Lowest 49 Levels} \\
    && \multicolumn{4}{c}{($E_u/k < 16900$ K)}  && \multicolumn{4}{c}{($E_u/k < 20000$ K)} \\
        && [3.6] & [4.5] & [5.8] & [8.0] && [3.6] & [4.5] & [5.8] & [8.0] \\
\tableline
$T$ (K) \vspace{0.1cm}\\
1000 && 99.9  & 99.9 & $>$99.9 &$>$99.9  &&$>$99.9  &$>$99.9  &$>$99.9  &$>$99.9  \\
2000 &&  84    & 97     & 99.5        &$>$99.0  && 96           & 99.2        & 99.8        &$>$99.9  \\
3000 && 44     & 89     & 96            &99.6        && 75           & 95            &98.0         &99.6\\
4000 && 21     & 78     & 89           &98.5         && 53           & 88           & 94            &98.5\vspace{0.1cm}\\

power index $s$ \vspace{0.1cm}\\
3.0 && 57 &92& 98.8 & $>$99.9 && 80 &96 &99.3 & $>$99.9 \\
3.8 && 65 &93 &99.2 & $>$99.9 && 84 &98 &99.5  & $>$99.9 \\
4.7 && 75 &94 &99.5 & $>$99.9 && 89 &97.5 &99.7  & $>$99.9 \\
6.0 && 87 &96 &99.8 & $>$99.9 && 95 &98.4 &99.9  & $>$99.9 \\ \tableline

\end{tabular}\\
\end{center}
\end{table*}

\begin{table*}
\begin{center}
\caption{Error of IRAC Color for LTE Calculations with the lowest 36/49 Levels\label{tbl_nLTE_Tmax2}}
\begin{tabular}{ccccccccc} \tableline \tableline
      && \multicolumn{3}{c}{Lowest 36 Levels}  && \multicolumn{3}{c}{Lowest 49 Levels} \\
    && \multicolumn{3}{c}{($E_u/k < 16900$ K)}  && \multicolumn{3}{c}{($E_u/k < 20000$ K)} \\
        && [3.6]--[4.5] & [4.5]--[5.8] & [5.8]--[8.0] && [3.6]--[4.5] & [4.5]--[5.8] & [5.8]--[8.0]  \\ 
\tableline
$T$ (K)\vspace{0.1cm}\\
1000 && $<$0.01 & $<$0.01 & $<$0.01 && $<$0.01 & $<$0.01 & $<$0.01 \\
2000 && 0.15  & 0.03 &  0.01                    && 0.03 & 0.01 & $<$0.01\\
3000 && 0.76   &0.09 & 0.04                     && 0.26 & 0.03 & 0.02\\
4000 && 1.43   &0.15 & 0.11                     && 0.56  &0.07 & 0.06 \vspace{0.1cm}\\

power index $s$\vspace{0.1cm}\\
3.0 && 0.51   & 0.08   & 0.02      && 0.20    &0.03    &0.01\\
3.8 && 0.38   & 0.07   & 0.01      && 0.15    &0.03    &0.01\\
4.7 && 0.25   & 0.06   & 0.01       && 0.10    &0.02    &$<$0.01\\
6.0 && 0.10   & 0.04   & $<$0.01 &&0.04    &0.02    &$<$0.01\\ \tableline
\end{tabular}\\
\end{center}
\end{table*}




\end{document}